\documentclass[12pt]{article}
\usepackage{epsf,rotating}
\newcommand{\mc}[3]{\multicolumn{#1}{#2}{#3}}
\begin{document}

\begin{center}
{\Large \bf Three years of Ulysses dust data: 1993 to 1995}
\end{center}


\bigskip

{ \bf
        H.~Kr\"uger\footnote{{\em Correspondence to:} 
krueger@galileo.mpi-hd.mpg.de}, 
        E.~Gr\"un$^1$,
        M.~Land\-graf$^1$,
        M.~Baguhl$^1$, 
        S.~Dermott$^2$,
        H.~Fechtig$^1$,
        B.~A.~Gustaf\-son$^2$,
        D.~P.~Hamilton$^3$, 
        M.~S.~Hanner$^4$,
        M.~Hor\'anyi$^5$,
        J.~Kissel$^1$,
        B.~A.~Lindblad$^6$,
        D.~Linkert$^1$,
        G.~Linkert$^1$,
        I.~Mann$^7$,
        J.~A.~M.~McDon\-nell$^8$,
        G.~E.~Morfill$^9$, 
        C.~Polanskey$^4$,
        G.~Schwehm$^{10}$,
        R.~Srama$^1$, and \\
\centerline{
        H.~A.~Zook$^{11}$
}
}

\bigskip

\small
\begin{tabular}{ll}
1)& Max-Planck-Institut f\"ur Kernphysik, 69029 Heidelberg, Germany\\
2)& University of Florida, Gainesville, FL\,32611, USA \\
3)& University of Maryland, College Park, MD\,20742-2421, USA\\
4)& Jet Propulsion Laboratory, Pasadena, California 91109, USA\\
5)& Laboratory for Atmospheric and Space Physics, Univ.
                 of Colorado, Boulder, \\ 
  & CO\,80309, USA\\
6)& Lund Observatory, 221 Lund, Sweden\\
7)& Max-Planck-Institut f\"ur Aeronomie, 37191 Katlenburg-Lindau, Germany\\ 
8)& University of Kent, Canterbury CT2 7NR, UK\\
9)& Max-Planck-Institut f\"ur Extraterrestrische Physik, 85748 Garching, 
                                                                   Germany\\ 
10)& ESTEC, 2200 AG Noordwijk, The Netherlands\\
11)& NASA Johnson Space Center, Houston, Texas 77058, USA\\
\end{tabular}

\normalsize

\bigskip

\begin{abstract}

The Ulysses spacecraft is orbiting the Sun on a highly inclined ellipse ($\rm 
i = 79^{\circ}$). After its Jupiter flyby in 1992 at a heliocentric 
distance of 5.4~AU, the spacecraft reapproached the inner solar system, 
flew over the Sun's south polar region in September 1994, crossed the 
ecliptic plane at a distance of 1.3~AU in March 1995, and flew over the 
Sun's north polar region in July 1995. We report on dust impact data 
obtained with the dust detector onboard Ulysses between January 1993 
and December 1995. We publish and analyse the complete data set of 509 
recorded impacts of dust particles with masses between $10^{-16}$~g to 
$10^{-7}$~g. Together with 968 dust impacts from launch until the end 
of 1992 published earlier (Gr\"un et al., 1995, {\em Planet. Space Sci}, 
Vol. 43, p.~971-999), information about 1477 particles detected with the 
Ulysses sensor between October 1990 and December 1995 is now available. 
The impact rate measured between 1993 and 1995 stayed relatively 
constant at about 0.4 impacts per day and varied by less than a factor of 
ten. Most of the impacts recorded outside about 3.5~AU are compatible with 
particles of interstellar origin. Two populations of interplanetary particles have 
been recognised: big micrometer-sized particles close to the ecliptic plane 
and  small sub-micrometer-sized particles at high ecliptic latitudes. The 
observed impact rate is compared with a model for the flux of interstellar 
dust particles which gives relatively good agreement with the observed impact rate. 
No change in the instrument's noise characteristics or 
degradation of the channeltron could be revealed during the three-year period.

\end{abstract}

\section{Introduction}

The Ulysses mission is exploring the solar system between 1 and 5.4 AU from 
the Sun over a wide range of ecliptic latitudes 
(-79$^{\circ}$ to +79$^{\circ}$).
Ulysses carries a multi-coincidence impact ionization detector, which is 
nearly identical to the dust detector flown onboard the Galileo spacecraft.
Detailed descriptions of the dust instruments onboard both spacecraft 
have been published by Gr\"un et al. (1992a,b, 1995a). Early results on 
interplanetary dust obtained from both missions and dust measurements
achieved by Ulysses in the environment of Jupiter 
have been reported (Gr\"un et al. 1992c,d, Gr\"un 1994). 
Unexpected intermittent streams of dust particles originating from 
the Jovian system and interstellar particles sweeping through the solar system 
were discovered by Ulysses (Gr\"un et al. 1993).
A detailed analysis of the complete Ulysses dust data set has led to the 
identification of ''small`` impacts that had been previously considered 
potential noise events (Baguhl et al. 1993). 

Data from the dust instruments onboard both spacecraft -- Ulysses and Galileo 
-- have been used in various ways: asteroids and comets have been investigated 
as dust sources (Riemann and Gr\"un 1992, Hamilton and Burns 1992, 
Gr\"un et al. 1994a, Mann et al. 1996a); consequences for the zodiacal light 
and other interplanetary meteoroid measurements have been considered (Mann 
et al. 1992; Mann and Gr\"un 1993, Mann et al. 1996b, Taylor et al. 1996); a 
detailed model of interplanetary meteoroid populations in the solar system has 
been developed (Divine 1993, Divine et al. 1993, Gr\"un and Staubach 1996, 
Gr\"un et al. 1997); the dust streams originating from the 
Jovian system have been  analysed (Hor\'anyi et al. 1993a,b, Hamilton 
and Burns 1993, Zook et al. 1996) and, finally, the properties of 
interstellar dust in the heliosphere  have been investigated (Gr\"un et al. 
1994b, Baguhl et al. 1995a,b, 1996, Landgraf and Gr\"un 1998). 

This is the fifth paper in a series dedicated to presenting both 
raw and reduced data obtained from the dust instruments onboard the 
Ulysses and Galileo spacecraft. Gr\"un et al. (1995a, hereafter Paper~I) 
describe the reduction process of Ulysses and Galileo dust data. 
Papers~II and III (Gr\"un et al. 1995b,c) 
present the data sets from the initial three and two years
of the Galileo and Ulysses missions, respectively. In the case of 
Ulysses the time period covered (Paper~III) was October 1990 to 
December 1992. In the current paper we extend the Ulysses data set 
from January 1993 until December 1995. In a companion paper 
(Kr\"uger et al. 1998, Paper~IV) we publish the Galileo data set for 
the same time period. The main data products are a table of the impact 
rate of all impacts determined from the particle accumulators and a table 
of both raw and reduced data of all dust impacts transmitted to Earth.
The information presented in these papers is similar to data which we 
are submitting to the various data archiving centers (Planetary 
Data System, NSSDC, Ulysses Data Center etc.). Electronic access to the 
data is also possible via the world wide web: http://galileo.mpi-hd.mpg.de.

This paper is organised similarly to Paper~III. Section~\ref{mission} 
gives a brief overview of the Ulysses mission and lists important mission 
events from 1993 to 1995. A description of the Ulysses dust 
data set for this period is given in Sect.~\ref{events}. 
Sections~\ref{analysis} and \ref{discussion} analyse and discuss the new 
data set.

\section{Mission and instrument operation} \label{mission}

\begin{figure}
\epsfxsize=8.5cm
\epsfbox{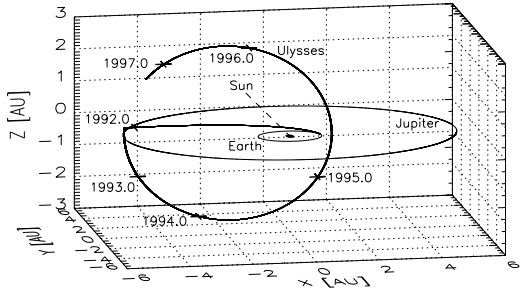}
        \caption{\label{trajectory}
Three-dimensional view of Ulysses' interplanetary trajectory from launch 
until the end of 1996 in ecliptic coordinates. The Sun is in the center. 
The orbits of Earth and Jupiter indicate the ecliptic plane. The initial
trajectory of Ulysses was in the ecliptic plane. During Jupiter flyby 
in early 1992 Ulysses was brought into an orbit with 79$^{\circ}$ inclination 
that sent the spacecraft close to the ecliptic poles.
Crosses mark the spacecraft position at the beginning of each year. Vernal
equinox is to the right (positive x axis).
}
\end{figure}

Ulysses was launched on 6 October 1990 and was brought onto a direct 
trajectory towards Jupiter. A swing-by maneuver at Jupiter 
on 8 February 1992 deflected the spacecraft into an 
orbit that is inclined by 79$^{\circ}$ to the ecliptic plane 
(Fig.~\ref{trajectory}). In October 1994 Ulysses passed the 
Sun's south polar region, then crossed the ecliptic plane again, this time at a 
perihelion distance 
of 1.3~AU on 12 March 1995 and flew over the Sun's north polar region in August 
1995. In April 1998 the spacecraft crossed the ecliptic 
plane at its aphelion at 5.4~AU. Approximate orbital elements for the 
Ulysses trajectory that include the whole out-of-ecliptic part of its orbit, 
are given in Paper~III. 

\begin{figure}
\epsfxsize=8.5cm
\epsfbox{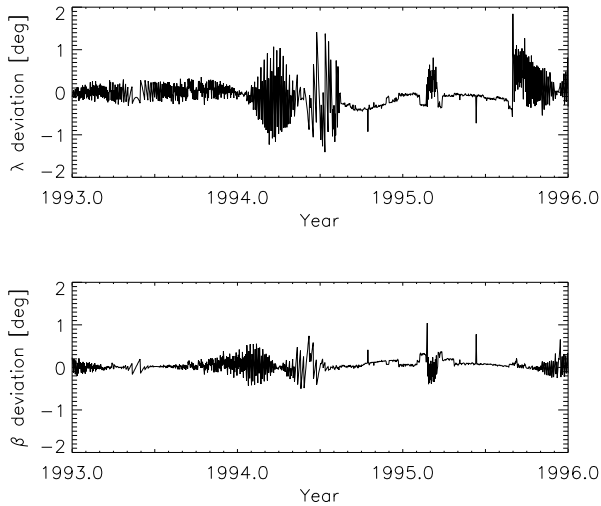}
        \caption{\label{pointing}
Spacecraft attitude: deviation of the antenna pointing direction 
(i.~e. negative spin axis) from the nominal Earth direction. The angles are 
given in ecliptic longitude (top) and latitude (bottom, equinox 1950.0).
}
\end{figure}

Ulysses is a spin-stabilized spacecraft with its spin axis pointing 
towards Earth. In Fig.~\ref{pointing} we show the deviation of the spin axis 
from the nominal Earth direction for the period 1993 to 1995 considered in 
this paper. Most of the time the axis pointing was within 
one degree of the Earth direction. This rather small deviation 
is negligible for the considerations in this paper. 
The Ulysses spacecraft and mission are explained in more detail by Wenzel 
et al. (1992). Details about the data transmission to Earth can also be 
found in Paper~III.

The dust detector onboard Ulysses (GRU) has a 140$^{\circ}$ wide field of view. 
The instrument is mounted nearly at right angles (85$^{\circ}$) to the antenna 
pointing direction 
(spacecraft spin axis). Therefore, the sensor is most sensitive to particles 
approaching from the plane perpendicular to the spacecraft-Earth direction. 
The rotation angle measures the sensor viewing direction at the time of a dust 
impact. During one spin revolution the rotation angle scans through a complete
circle of 360$^{\circ}$. The 0$^{\circ}$ rotation angle is defined to be the 
direction closest to ecliptic north. At high ecliptic latitudes, however, the 
sensor pointing at 0$^{\circ}$ rotation angle significantly deviates from the 
actual north direction. During the passages over the Sun's polar regions the sensor 
always scans through a plane tilted by about $30^{\circ}$ from the ecliptic and all 
rotation angles lie close to the ecliptic plane 
(cf. Fig.~4 in Gr\"un et al. 1997). A 
sketch of the viewing geometry can be found in Gr\"un et al. (1993).
 
Table~\ref{event_table} lists significant mission and dust instrument 
events from 1993 to 1995. Earlier events are only listed if especially significant.
A comprehensive list of events from launch until the end of 1992 is given in 
Paper~III. During the early phases of the mission 
the in-orbit noise characteristics of the instrument were investigated 
(Paper~III). This led to a relatively noise-free configuration (hereafter called
nominal configuration) for the instrument after 10 February 1992:
channeltron voltage 1140~V (HV~=~3); event definition status such that either the 
channeltron or the ion-collector channel can, independent of each other,
start a measurement cycle 
(EVD~=~C,~I); detection thresholds for ion-collector, channeltron and 
electron-channel set to the lowest levels and the detection threshold for 
the entrance grid set to the first digital step (SSEN~=~0,~0,~0,~1). See 
Paper~I for a description of these terms.

The operational configuration of the dust instrument 
was changed several times during noise tests: 
starting from the nominal configuration described above, all tests have been 
performed with the same instrument settings. During noise tests, the 
following changes 
of the instrument configuration were applied at one-hour intervals:
a) set the event definition status such that the channelton, the ion 
collector and the electron-channel can initiate a measurement cycle 
(EVD~=~C,~I,~E);
b) set the thresholds for all channels to their lowest levels 
(SSEN~=~0,~0,~0,~0);
c) reset the event definition status to its nominal configuration (EVD~=~C,~I);
d) increase the channeltron high voltage by one digital step (HV~=~4);
e) reset the channeltron high voltage and the detection thresholds to their
nominal settings (HV~=~3, SSEN~=~0,~0,~0,~1).
After step e) the instrument is back in its nominal configuration. 
No detectable change in the noise behaviour was revealed by the noise 
tests during the three years from 1993 to 1995.

Spacecraft anomalies occurred five times between 1993 and 1995, and all 
scientific instruments onboard were switched off automatically for about one day 
during so-called DNELs (Disconnect all Non-Essential Loads). After each DNEL, the 
dust instrument was switched on again and was configured to its nominal 
operational mode. 

The dust instrument has two heaters to allow for a relatively stable operating 
temperature within the sensor. By heating one of the two or both heaters, three 
different heating power levels can be achieved (400~mW, 800~mW or 1,200~mW).
One or both heaters were switched on most of the time, except close to 
the Sun between 5 February  and 12 June 1995 when both were switched off. 
The heaters remained switched on 
during the DNELs. Table ~\ref{event_table} lists the total heating powers provided 
by the heaters. The temperature of the dust sensor was between 
-20$^{\circ}\,\rm C$ and +15$^{\circ}\,\rm C$. 

A timing error in the instrument electronics led to wrong spacecraft sector 
information for about 20\% of the events in the data set published 
earlier (cf. Paper~III, indicated by ROT\,=\,999 in Table 4). 
The error was corrected by a reprogramming of the instrument on 30 April 1993.

After launch of Ulysses the sounder of the URAP instrument (Stone et al.\ 1992) 
emerged as a significant and unexpected noise source for the dust sensor.
When the sounder was switched on after launch, sounder interference caused about 
20\% dead time (Paper~III). With the sounder now usually being operated at a 
lower rate of about 2~min 
at 2~hour intervals the dead time is reduced to about 2\%. In a detailed 
investigation of the noise behavior of the Ulysses dust instrument Baguhl et al. 
(1993) showed that the noise rate measured during periods of sounder operation 
was correlated with the distance to, and the position of, the Sun with 
respect to 
the sensor-viewing direction: most noise events are triggered when the Sun 
shines directly into the sensor. On the other hand, when the spacecraft was at 
large heliocentric distances in 1993, the noise rate was 
extremely low, even during sounder operation periods.

\begin{figure}
\epsfxsize=8.5cm
\epsfbox{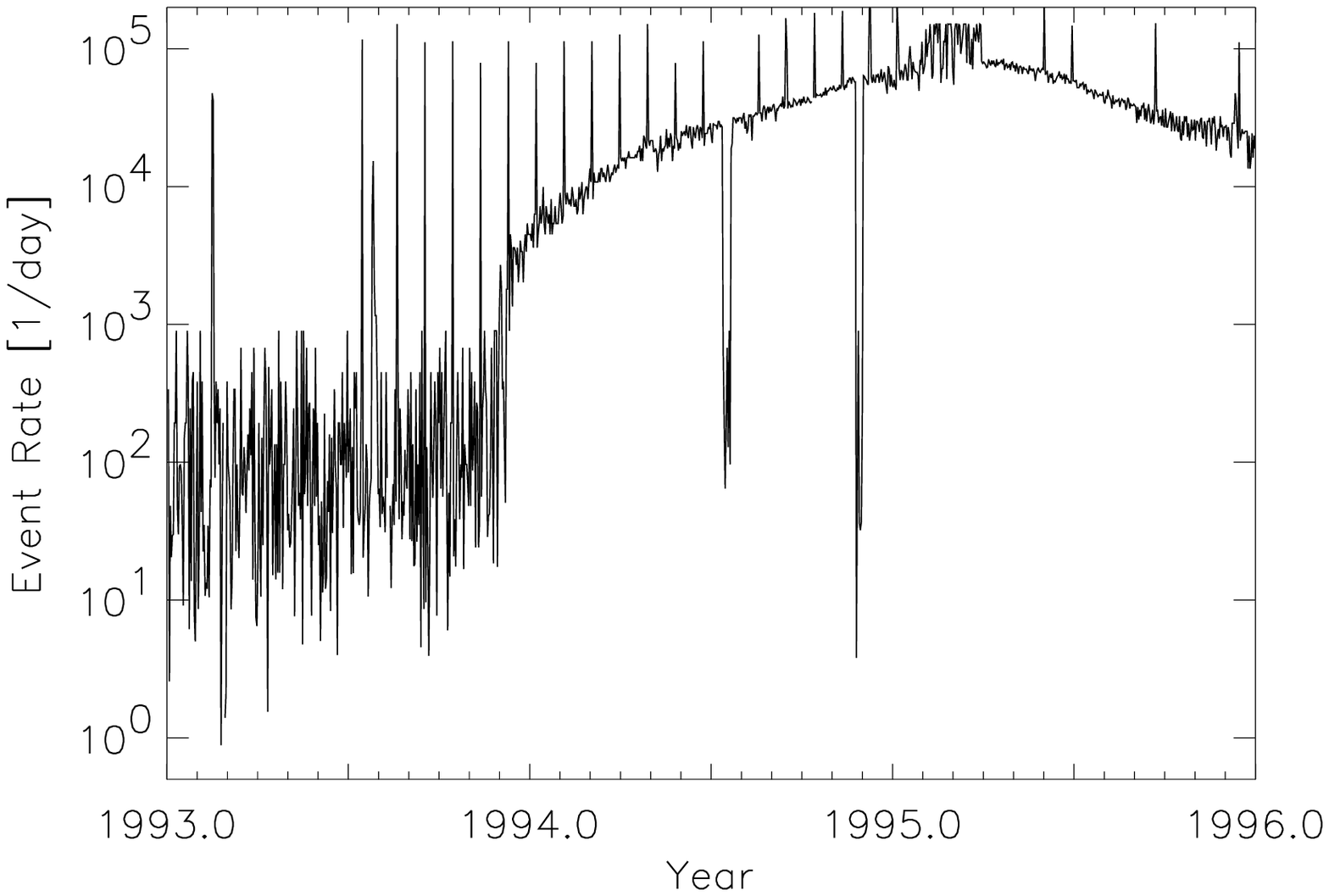} 
\epsfxsize=8.5cm
\epsfbox{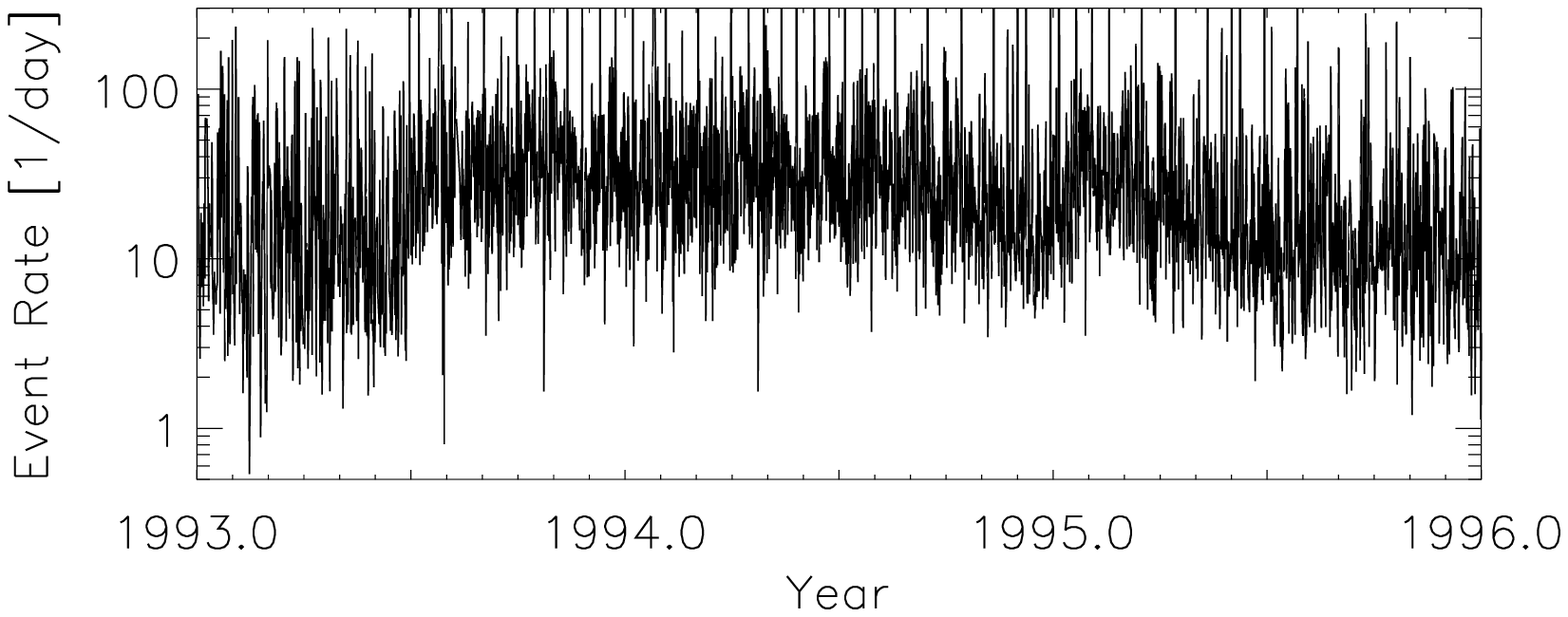}
        \caption{\label{noiserate}
Noise rate (class~0 events) detected with the dust instrument. 
Upper panel: Daily 
maxima in the noise rate (determined from the AC01 accumulator). The 
sounder was operated for only 2\,\% of the total time, and the daily 
maxima are dominated by sounder noise. Sharp spikes 
are caused by periodic noise tests and short periods of 
reconfiguration after DNELs (cf. Table~\ref{event_table}). From 12 
to 22 Jul 1994 and 24 Nov to 1 Dec 1994 the sounder was not operated 
which reduced the maximum noise by several orders of magnitude. 
Lower panel: Noise rate detected during quiet intervals when the 
sounder was switched off, which was the case about 
98\,\% of the time (one-day average 
calculated from the number of AC01 events for which the complete 
information has been transmitted to Earth).
}
\end{figure}

In Fig.~\ref{noiserate} we show the noise rate for the 1993 to 1995 period. 
The upper panel shows the daily maxima, which are dominated by 
interference with the sounder. Since the sounder was operated for periods of
only 2~min with quiet intervals of about 2~hours, such high noise rates 
prevailed only during about 2\% of the time, with the remaining 98\% 
being free of sounder noise. In 1994 the maxima in the noise rate 
induced by the sounder began to increase significantly when Ulysses 
approached the inner solar system. The highest sounder noise rates 
occurred around perihelion passage in March 1995. From
12 to 22 July 1994 and 24 November to 1 December 1994 the sounder was
switched off and the noise level dropped to about 10 events per day.
The noise during quiet times when 
the sounder was switched off is shown in the lower panel of 
Fig.~\ref{noiserate}. The average was about 20 events per day which 
shows that the dust instrument was not affected by dead time 
caused by random noise events during 98\% of the time.

\section{Impact events} \label{events}

Impact events are classified into four classes and six ion charge amplitude 
ranges which lead to 24 individual categories. In addition, the instrument
has 24 accumulators with one accumulator belonging to one individual 
category. Class~3, our
highest class, are real dust impacts and class~0 are noise events. 
Depending upon the noise of the charge measurements, classes~1 and 2 
can be true dust impacts or noise events. This classification 
scheme for impact events has been described in Paper~I and this 
scheme is still valid for the Ulysses dust instrument. In contrast to 
the Galileo dust instrument which had to be reprogrammed because of 
the low data transmission capabilities of the Galileo spacecraft, no such 
reprogramming was necessary for Ulysses. Most of the data processing 
for Ulysses is done on the ground.

Between 1 January 1993 and 31 December 1995 the complete data (sensor 
orientation, charge amplitudes, charge rise times, etc.) of 72,809 events 
including 509 dust impacts were transmitted to Earth. Table~2 
lists the number of all dust impacts counted with the 24 accumulators of 
the dust instrument. `AC{\em xy}' refers to class number `$x$' and amplitude
range `$y$' (for a detailed description of the accumulator categories
see Paper~I). As discussed in the previous section, most noise events were 
recorded during the short time 
periods when either the sounder of the URAP instrument was operating (cf. 
Paper~III) or the dust instrument was configured to its high sensitive state 
for noise tests, or both. During these periods many events were only counted 
by one of the 24 accumulators
because their full information was overwritten before the data could be
transmitted to Earth (see bottom of Table~2). Since the dust impact
rate was low during times surrounding these periods, it is expected
that only a few true dust impacts were lost. 

Two particles in AC32 and AC33 (day 93-154, 21:01\,h and 93-319, 15:31\,h), 
respectively, were detected in a gap of several hours when no data could be 
transmitted to Earth, and their full information had to be taken from the 
instrument memory (FN7, cf. Gr\"un et al. 1992a). Before and after these 
gaps the sounder was operated for about 2 min at 2 hour intervals and we 
assume that it was operated with the same frequency in the gap. 
Due to these long quiet intervals between sounder operation it is rather 
unlikely that the two events occurred when the sounder was active.
Even in such a case these are still very likely true dust impacts because the 
sounder noise usually affects only the lowest ion amplitude range (AR1).

All 509 dust impacts detected between 1993 and 1995 for which the complete 
information exists are listed in Table~3. Dust particles are identified by their 
sequence number and their impact time (first two columns). The event category -- 
class (CLN) and amplitude range (AR) -- are given in the third and fourth 
columns. Raw data as transmitted to Earth 
are shown in the next columns: sector value (SEC) which is the
spacecraft spin orientation at the time of impact,
impact charge numbers (IA, EA, CA) and rise times (IT, ET), time
difference and coincidence of electron and ion signals (EIT, EIC),
coincidence of ion and channeltron signal (IIC), charge reading at
the entrance grid (PA) and time (PET) between this signal and
the impact. Then the instrument configuration is given: event
definition (EVD), charge sensing thresholds (ICP, ECP, CCP, PCP) and
channeltron high voltage step (HV). Compare Paper~I for further
explanation of the instrument parameters. 

The next four columns in Table~3 give information about Ulysses' orbit: 
heliocentric distance (R), ecliptic longitude and latitude (LON, LAT) 
and distance from Jupiter ($\rm D_{Jup}$). The next column gives the 
rotation angle (ROT) as described in Sect.~\ref{mission}. 
Whenever this value is unknown, ROT is arbitrarily set to
999. This occurs 14 times. Then follows the pointing direction of the 
dust instrument at 
the time of particle impact in ecliptic longitude and latitude 
($\rm S_{LON}$, $\rm S_{LAT}$).
When ROT is not valid $\rm S_{LON}$ and $\rm S_{LAT}$ are useless and are 
also set to 999. Mean
impact velocity (V) and velocity error factor (VEF, i.e. multiply or
divide stated velocity by VEF to obtain upper or lower limits) as well 
as mean particle mass (M) and mass error factor (MEF) are given in the 
last columns. For VEF $> 6$, both velocity and mass values should be
discarded. This occurs for 48 impacts. No intrinsic dust charge values 
are given (see Svestka et al. 1996 for a detailed analysis).

\section{Analysis} \label{analysis}

\begin{figure}
\epsfxsize=8.0cm
\epsfbox{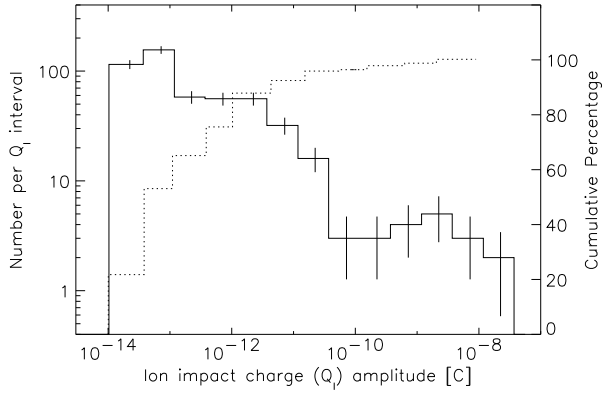}
        \caption{\label{nqi}
Distribution of the impact charge amplitude $\rm Q_I$ for all particles 
detected between 1993 and 1995. The solid line
indicates the number of impacts per charge interval, and the 
dotted line shows the cumulative percentage. Vertical bars
indicate the $\rm \sqrt{n}$ statistical error. 
}
\end{figure}

The positive impact charge measured on the ion collector, $\rm Q_I$, is 
the most important impact parameter determined by the dust instrument 
because of its relative insensitivity to noise. In Fig.~\ref{nqi} we show the 
distribution of $\rm Q_I$ for all dust particles detected between 1993 and 
1995. Ion impact charges have been detected over the entire range of six 
orders of magnitude in impact charge that can be measured by the dust instrument. 
About 1\% of all impacts are close to the saturation limit of $\rm Q_I \sim 
10^{-8}\,C$ and may thus 
constitute lower limits of the actual impact charges. The impact charge 
distribution is reminiscent of three individual particle populations: 
small particles with impact charges $\rm Q_I<10^{-13}\,C$ (AR1), intermediate
particles with $ \rm 10^{-13}\,C \leq Q_I \leq 3\times 10^{-11}\,C $ 
(AR2 and AR3) and big 
particles with $\rm Q_I > 3\times 10^{-11}\,C$ (AR4 to AR6). The intermediate 
particles are mostly of interstellar origin and the big particles are 
interplanetary particles detected close to the ecliptic plane 
(Gr\"un et al., 1997, see also Sect.~\ref{discussion}). The small particles 
occur mostly over the Sun's polar regions and are attributed to a population of
interplanetary $\beta$-meteoroids at high ecliptic latitudes (Baguhl et al., 
1995b, Hamilton et al., 1996).

\begin{figure}
\epsfxsize=8.0cm
\epsfbox{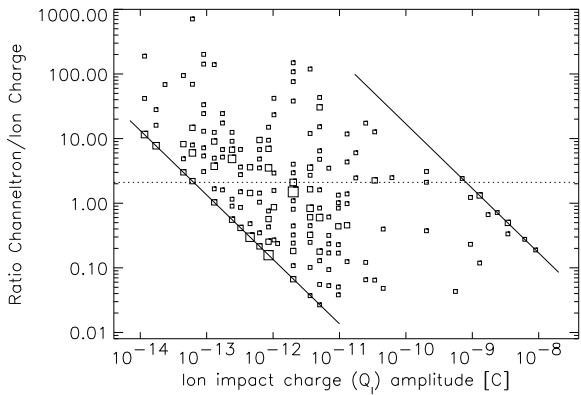}
        \caption{\label{qiqc}
Channeltron amplification factor $\rm A = Q_C/Q_I$
as a function of impact charge $\rm Q_I$ for all particles 
detected between 1993 and 1995. The solid lines denote the sensitivity
threshold (lower left) and the saturation limit (upper right) of the channeltron. Squares
indicate dust particle impacts. The area of each square is proportional to 
the number of events included (the scaling of the squares differs 
from that used in Paper~III). The dotted horizontal line shows the mean value 
of the channeltron amplification A\,=\,2.1 
for ion impact charges $\rm 10^{-12}~C < Q_I < 10^{-11}~C$.
}
\end{figure}

The ratio of the channeltron charge $\rm Q_C$ and the ion collector
charge $\rm Q_I$ is a measure of the channeltron amplification A.
The channeltron amplification is an important parameter for dust 
impact identification (Paper~I).
In Fig.~\ref{qiqc} we show the charge ratio $\rm Q_C/Q_I$ as a function of 
$\rm Q_I$ for the nominal high voltage of 1140\,V (HV\,=\,3). The mean 
amplification determined from particles with $\rm 10^{-12}{\rm\,C} \le Q_I
\le 10^{-11}{\rm\,C}$ is $\rm A \simeq 2.1$. This is very close to 
the value obtained in Paper~III for the first two years of the mission
($\rm A \simeq 2.2$). Therefore, the channeltron does not 
show any detectable aging during the more than five years of 
the Ulysses mission.

\begin{figure}
\epsfxsize=8.5cm
\epsfbox{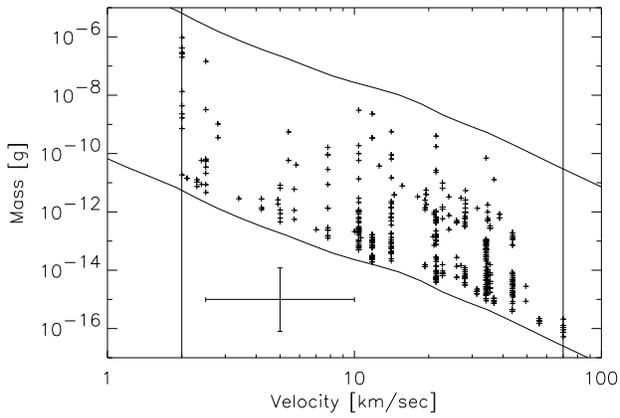}
        \caption{\label{mass_speed}
Masses and impact velocities of all impacts recorded with the Ulysses sensor 
between 1993 and 1995. The lower and upper solid lines indicate the threshold and
the saturation limit of the detector, respectively, and the vertical lines indicate the 
calibrated velocity range. A sample error bar is shown that indicates
a factor of 2 uncertainty for the velocity and a factor of 10 for the mass determination.
}
\end{figure}

Figure~\ref{mass_speed} shows the masses and velocities of
all dust particles detected between 1993 and 1995. As in the
earlier period (1990 to 1992, Paper~III) velocities occur over
the entire calibrated range from 2 to 70 km/s. The masses
vary over 10 orders of magnitude from $\rm 10^{-6}\,g$ to
$\rm 10^{-16}\,g$. The mean errors are a factor of 2 for the
velocity and a factor of 10 for the mass. The clustering
of the velocity values is due to discrete steps in the rise
time measurement but this quantization is much smaller than the
velocity uncertainty. 
For many particles in the lowest two amplitude ranges (AR1 and
AR2) the velocity had to be computed from the ion charge signal
alone which leads to the striping in the lower mass range in 
Fig.~\ref{mass_speed} (most prominent above 10 km/s). In the
higher amplitude ranges the velocity could normally be calculated
from both the target and the ion charge signal, resulting in a 
more continuous distribution in the mass-velocity plane. Impact 
velocities below about 3 km/s should be treated with caution
because anomalous impacts onto the sensor grids or structures 
other than the target generally lead to prolonged rise times and 
hence to unnaturally low impact velocities. 

\section{Discussion} \label{discussion}

\begin{figure}
\epsfxsize=8.5cm 
\epsfbox{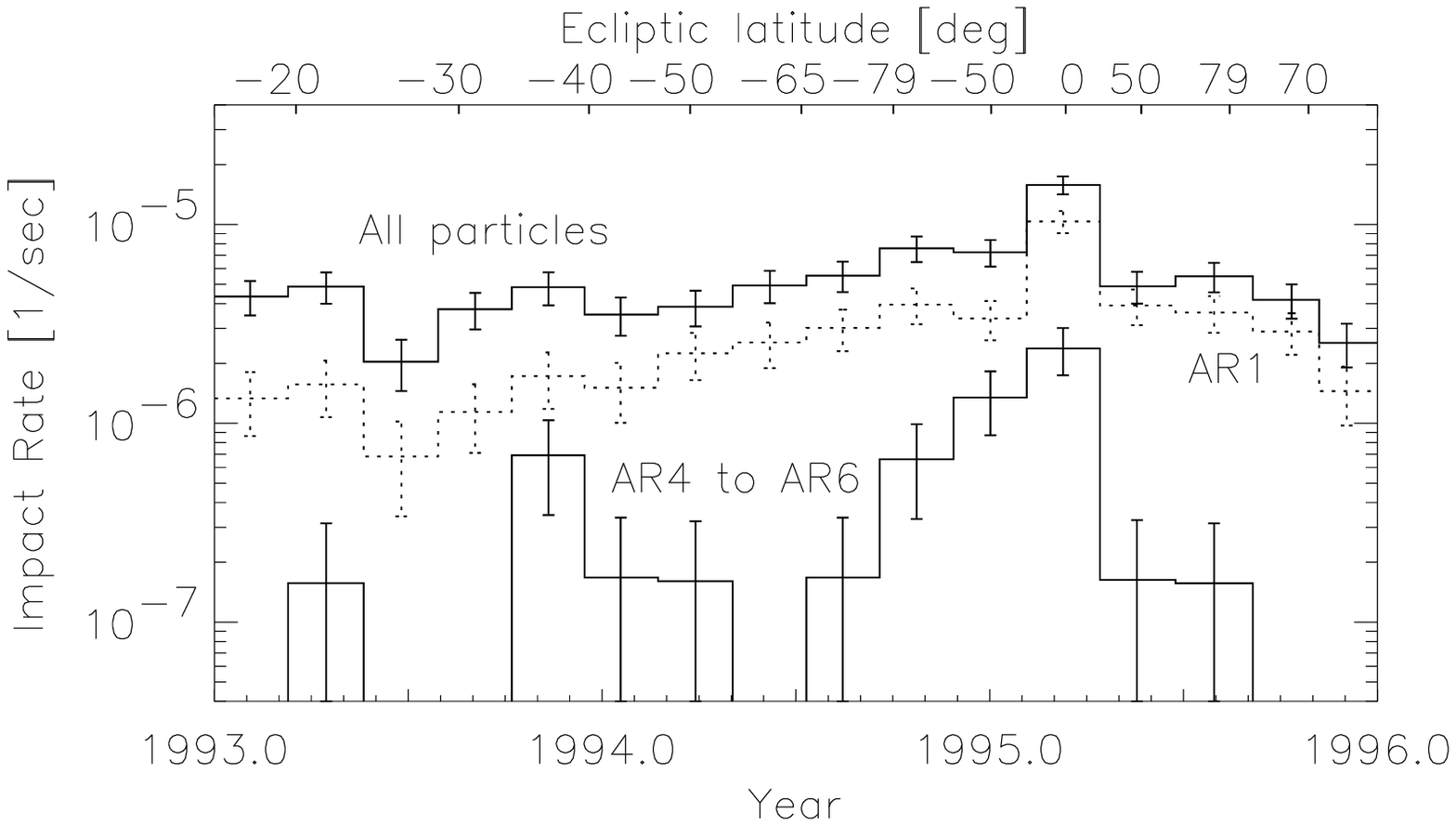} 
\epsfxsize=8.5cm
\epsfbox{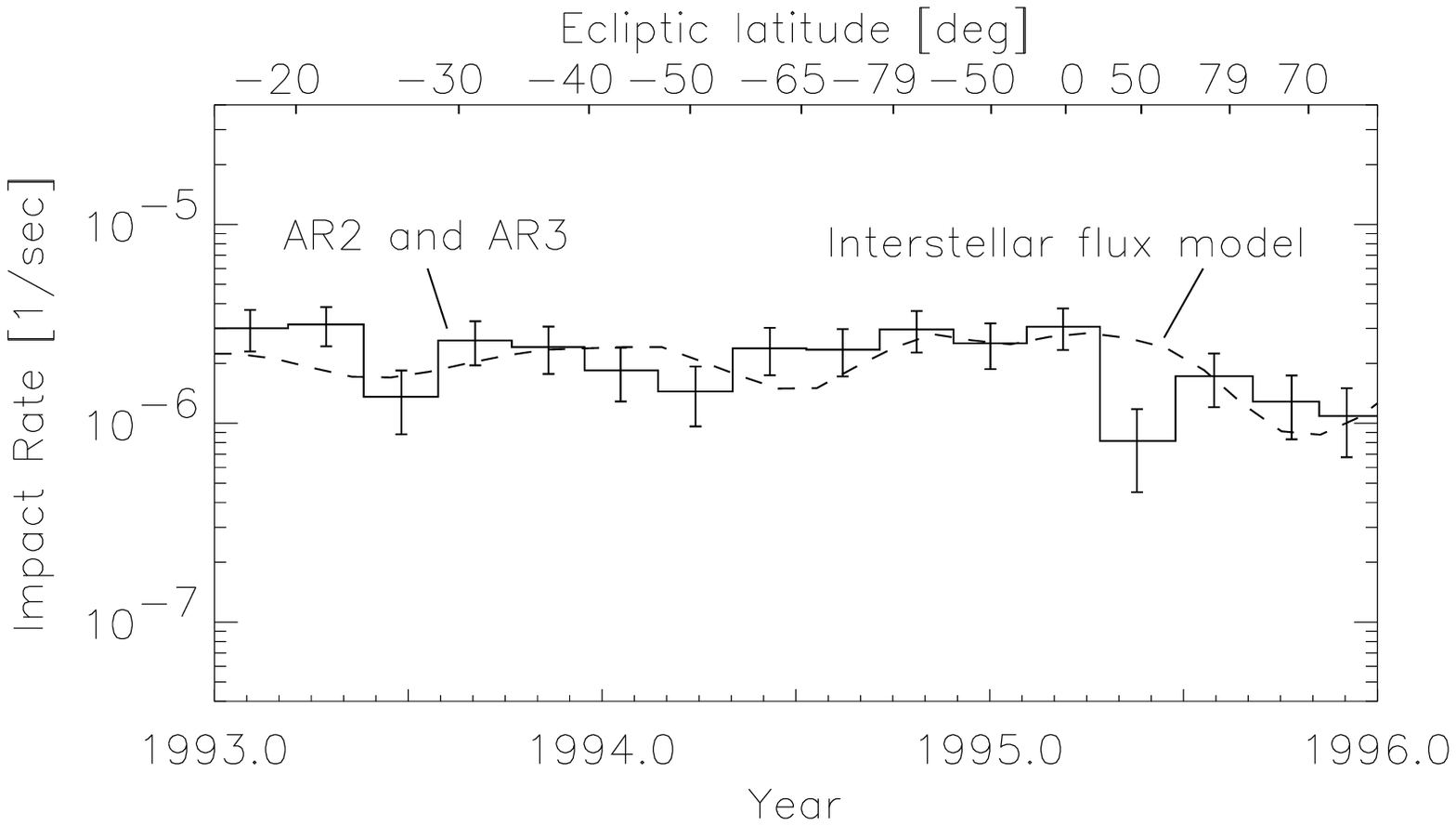}
        \caption{\label{rate}
Impact rate of dust particles detected with the Ulysses dust sensor 
as a function of time with the ecliptic latitude of the spacecraft 
indicated at the top. Upper panel: the upper solid line shows the total 
impact rate, the dotted line the impact rate of small particles 
(AR1) and the lower solid line the rate of big particles (AR4 to AR6). 
Note that a rate of about $\rm 1.8 \times 10^{-7}$ impacts per second 
is caused by a single dust impact in the averaging interval of about 
70 days.  Lower panel: intermediate particles (AR2 and AR3, 
solid line). A model for the rate of interstellar particles assuming 
a constant flux is superimposed as a dashed line. 
}
\end{figure}

Most of the time from January 1993 to December 1995 Ulysses 
was at high ecliptic latitudes far away from the ecliptic plane.
The dust impact rate detected by the Ulysses dust sensor in 
this period  is displayed in Fig.~\ref{rate}. 
The highest overall impact rate was detected around 
the ecliptic plane crossing which occurred on 12 March 1995 
(at 1.3~AU from the Sun). The maximum impact rate 
of particles in the three highest ion amplitude ranges 
(AR4 to AR6) coincides with the highest overall impact rate
(upper panel of Fig.~\ref{rate}). 
These impacts are attributed to interplanetary particles 
on low inclination orbits (Gr\"un et al. 1997). These are the 
impacts with $\rm  Q_I > 10^{-10}~C$ shown in Fig.~\ref{nqi}. 
The impact rate of particles in the lowest amplitude range (AR1)
increased gradually since mid 1993, reached its maximum at the ecliptic 
plane crossing in March 1995 and decreased later. Although the impact rate 
reached its maximum during a short time around the ecliptic plane 
crossing, the majority of these small particles has been detected 
during the much longer time interval when Ulysses was at high ecliptic 
latitudes. They are attributed to a population of small 
interplanetary particles on escape trajectories from the solar 
system (Baguhl et al. 1995, Hamilton et al. 1996).

\begin{figure}
\epsfxsize=8.5cm
\epsfbox{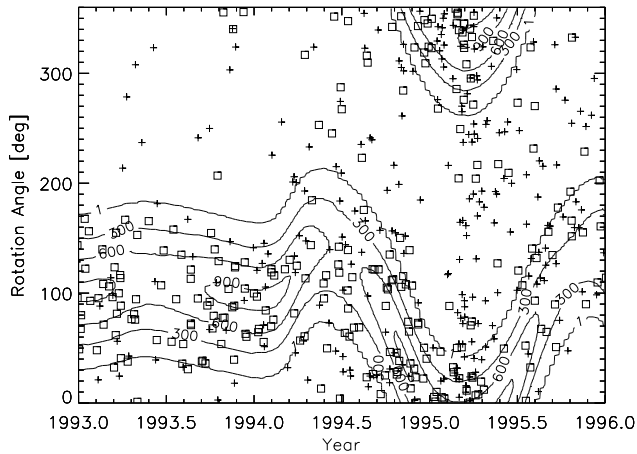}
        \caption{\label{rot_angle} 
Rotation angle vs. time for all particles detected between 1993 and 1995 for 
which complete information is available. Plus signs indicate particles 
with impact charge $\rm Q_I < 8 \times 10^{-14}$~C, squares those with
$\rm Q_I \geq 8 \times 10^{-14}$~C. The contour lines show the sensitive 
area of the dust sensor for particles approaching from the interstellar 
upstream direction (levels of 1, 300, 600 and $\rm 900\,cm^{2}$ detector area 
are shown). Most of the larger $\rm Q_I$ particles came from directions
consistent with the interstellar upstream direction. The biggest 
particles in the highest amplitude ranges which are of interplanetary 
origin are not shown separately here because only very few of them 
were detected (cf. Table~2) and they cannot be separated from interstellar
particles by directional arguments. Passage over the Sun's 
south polar region occurred in October 1994, ecliptic plane crossing 
in March 1995, and passage over the Sun's north polar region in
August 1995.
}
\end{figure}

The impact rate of particles in the intermediate ion amplitude ranges 
AR2 and AR3
was relatively constant during the 1993 to 1995 period. It dominated 
the overall impact rate until early 1994, i.e. outside 
about 3.5~AU from the Sun. Impacts in these ion amplitude ranges are 
mostly due to interstellar particles (see also Fig.~\ref{rot_angle} 
and Gr\"un et al. 1993, Baguhl et al. 1995).
Figure~\ref{rate} also shows the expected impact rate of interstellar 
particles assuming that they approach from the direction of interstellar 
helium (Witte et al. 1996) and that they move on straight trajectories with 
a relative velocity of $26\ {\rm km}\ {\rm s}^{-1}$ through the solar 
system. This assumption means dynamically that radiation pressure
cancels gravity for these particles ($\beta=1$) and that their 
Larmor radii are large compared with the dimension of the solar system. 
Both assumptions are reasonable for particles with masses between 
$10^{-13}\ {\rm g}$ and $10^{-12}\ {\rm g}$ which is the dominant 
size range measured for interstellar particles (e.~g. Gr\"un et al. 
1997). The dust particle flux is independent of heliocentric distance 
in this simple model, which gives relatively good agreement with the
observed impact rate. The variation predicted 
by the model is caused by changes in the instrument's viewing 
direction with respect to the approach direction of the particles 
and changes in the relative velocity between the spacecraft and 
the particles. 
After the ecliptic plane crossing the rate of interstellar particles 
expected from the model is significantly higher than the one observed. 
A detailed dynamical model (Landgraf 1998) for the motion of 
interstellar particles 
in the interplanetary magnetic field gives 
better agreement with the observed impact rate, especially 
after 1996.

The sensor orientation at the time of a particle
impact (rotation angle) is shown in Fig.~\ref{rot_angle}. 
The detector's sensitive area for particles approaching from the
interstellar direction is superimposed. The
bigger particles (squares, impact charge $\rm Q_I \geq 8 \times 
10^{-14}~C$) are clearly concentrated towards the interstellar 
direction. They have been detected with a relatively constant 
rate during the whole three-year period (Fig.~\ref{rate}). 
Only a few small particles
(crosses, impact charges $\rm Q_I \leq 8 \times 10^{-14}~C$) have been
detected and they cluster above the Sun's polar regions.
The particles with the highest ion amplitude 
ranges (AR4 to AR6) are not distinguished in this diagram because 
they cannot be separated from interstellar particles by directional 
arguments. They have to be distinguished by other means
(e.~g. mass and velocity).  Furthermore, their total number is so small 
that they constitute only a small 'contamination' of the interstellar 
particles in Fig.~\ref{rot_angle}. In the ecliptic plane at 1.3~AU, 
however, interplanetary particle flux dominates over interstellar flux 
by a factor of about 3 (in number).

Streams of tiny dust particles originating from the Jovian system 
have been first detected with Ulysses and later with Galileo out to 
2~AU from Jupiter (Gr\"un et al., 1993, 1996). During the time period 
from 1993 to 1995 considered here, Ulysses was more than 2.5~AU away 
from Jupiter which makes a significant contribution of Jovian dust 
stream particles in the present data set very unlikely.
Analysis of the particles' trajectories and their interaction 
with the interplanetary magnetic field (Zook et al., 1996) showed 
that the velocities 
are about 300 km/s and the particles are about 10~nm in 
size. These values are far beyond the calibrated range of the dust 
sensors given above and, in principle, one cannot exclude that the 
data set presented in this paper contains at least a fraction of
particles for which the masses and velocities inferred from our 
calibration are wrong. Recent analyses of the interplanetary 
(Gr\"un and Staubach, 1997) and interstellar populations 
(Landgraf, 1998), however, are 
consistent with the calibrated velocities and masses of the particles 
in the new 1993 to 1995 data set. Given 
the present knowledge about the particles' dynamics,
the velocities and masses stated in Papers~II and III 
for the stream particles are not correct.
Future investigations will address possible contributions of 
particles beyond the calibrated range of the dust instruments 
in the complete Ulysses and Galileo data sets, not only close 
to Jupiter where the dust streams have been detected. 

\hspace{1cm}

{\bf Acknowledgments.}
This work has been supported by the Deutsche Agentur f\"ur 
Raumfahrtangelegenheiten (DARA).

\section*{References}

{\small

{\bf Baguhl, M., Gr\"un, E., Linkert, D., Linkert, G.\ and Siddique, N.,}
Identification of 'small' dust impacts in the Ulysses dust
detector data. {\em  Planet.\ Space Sci. }{\bf 41}, 1085-1098, 1993

{\bf Baguhl, M., Gr\"un, E., Hamilton, D.P., Linkert, G., Riemann, R.,
Staubach, P.\ and Zook H.,} The flux of interstellar dust observed by
Ulysses and Galileo. {\em Space Sci.\ Rev. }{\bf 72}, 471-476, 1995a

{\bf Baguhl, M., Hamilton, D.P., Gr\"un, E., Dermott, S., Fechtig, H., 
Hanner, M. S., Kissel, J., Lindblad, B.-A., Linkert, D., Linkert, G., Mann, I.,
McDonnell, J. A. M., Morfill, G. E., Polanskey, C., Riemann, R., 
Schwehm, G., Staubach, P. and Zook, H.,} Dust measurements at high
ecliptic latitudes. {\em Science} {\bf 268}, 1016-1019, 1995b

{\bf Baguhl, M., Gr\"un, E. and Landgraf, M.,} In situ measurements of 
interstellar dust with the Ulysses and Galileo spaceprobes. 
{\em Space Sci Rev.}, {\bf 78}, 165-172, 1996

{\bf Divine, N.\,} Five populations of interplanetary meteoroids.
{\em J.~Geophys.~Res. }{\bf 98}, 17029-17048, 1993

{\bf Divine, N.\, Gr\"un, E. and Staubach, P.,} Modelling the meteoroid 
distributions in interplanetary space and near Earth. 
{\em Proc. First European Conference on Space Debris}, Darmstadt, ed. W. Flury, 
ESA SD-01, 245-250, 1993 

{\bf Gr\"un, E.,} Dust measurements in the outer solar system.
In: {\em Asteroids, Comets, Meteors 1993}, eds. A Melani, M. Di Martino, 
A. Cellino, Kluwer Acad. Publ., 367-380, 1994

{\bf Gr\"un, E., Fechtig, H., Hanner, M.S., Kissel, J., Lindblad, B-A.,
Linkert, D., Linkert, G., Morfill, G.E.\ and Zook, H.A.,}
The Galileo dust detector. {\em Space Sci.\ Rev. }{\bf 60}, 317-340, 1992a

{\bf Gr\"un, E., Fechtig, H., Giese, R.H., Kissel, J., Linkert, D.,
Maas, D., McDonnell, J.A.M., Morfill, G.E., Schwehm, G.\ and
Zook, H.A.,} The Ulysses dust experiment.
{\em Astron.\ Astrophys.\ Suppl.\ Ser. }{\bf 92}, 411-423, 1992b

{\bf Gr\"un, E., Baguhl, M., Fechtig, H., Hanner, M.S., Kissel, J.,
Lindblad, B.-A., Linkert, D., Linkert, G., Mann, I., McDonnell, J.A.M.,
Morfill, G.E., Polanskey, C., Riemann, R., Schwehm, G., Siddique, N. and
Zook, H.A.,} Galileo and Ulysses dust measurements: From Venus
to Jupiter. {\em Geophys.\ Res.\ Letters }{\bf 19}, 1311-1314, 1992c

{\bf Gr\"un, E., Zook, H.A., Baguhl, M., Fechtig, H., Hanner, M.S., 
Kissel, J., Lindblad, B.-A., Linkert, D., Linkert, G., Mann, I., 
McDonnell, J.A.M., Morfill, G.E., Polanskey, C., Riemann, R., Schwehm, G. and
Siddique N., } Ulysses dust measurements near Jupiter. {\em
Science }{\bf 257}, 1550-1552, 1992d

{\bf Gr\"un, E., Zook, H.A., Baguhl, M., Balogh, A., Bame, S.J.,
Fechtig, H., Forsyth, R., Hanner, M.S., Horanyi, M., Kissel, J.,
Lindblad, B.-A., Linkert, D., Linkert, G., Mann, I., McDonnell, J.A.M.,
Morfill, G.E., Phillips, J.L., Polanskey, C., Schwehm, G., Siddique, N.,
Staubach, P., Svestka, J. and Taylor, A., } Discovery of jovian
dust streams and interstellar grains by the Ulysses spacecraft. {\em
Nature }{\bf 362}, 428-430, 1993

{\bf Gr\"un, E., Hamilton, D.P., Baguhl, M., Riemann, R., Horanyi, M.\ and
Polan\-skey, C.,} Dust streams from comet Shoemaker-Levy 9? {\em
Geophys.\ Res.\ Let. }{\bf 21}, 1035-1038, 1994a

{\bf Gr\"un, E., Gustafson, B., Mann, I., Baguhl, M., Morfill, G.E.,
Staubach, P., Taylor, A. and Zook, H.A., } Interstellar dust in the
heliosphere. {\em Astron. Astrophys. }{\bf 286}, 915-924, 1994b

{\bf Gr\"un, E., Baguhl, M., Fechtig, H., Hamilton, D.P., Kissel, J.,
Linkert, D., Linkert, G.\ and Riemann, R.,}
Reduction of Galileo and Ulysses dust data.
{\em Planet. Space Sci.} {\bf 43}, 941-951, 1995a (Paper I)

{\bf Gr\"un, E., Baguhl, M., Divine, N., Fechtig, H., Hamilton, D. P.,
Hanner, M. S., Kissel, J., Lindblad, B.-A., Linkert, D., Linkert, G., 
Mann, I., McDonnell, J. A. M., Morfill, G. E., Polanskey, C., Riemann, R.,
Schwehm, G., Siddique, N., Staubach P. and Zook, H. A.,}
Three years of Galileo dust data. {\em Planet. Space Sci.}, {\bf 43}, 
953-969, 1995b (Paper II)

{\bf Gr\"un, E., Baguhl, M., Divine, N., Fechtig, H., Hamilton, D.P.,
Hanner, M.S., Kissel, J., Lindblad, B.-A., Linkert, D., Linkert, G.,
Mann, I., McDonnell, J.A.M., Morfill, G.E., Polanskey, C.,
Riemann, R., Schwehm, G., Siddique, N., Staubach, P.\ and Zook,
H.A., } Two years of Ulysses dust data, {\em Planet. Space Sci.} 
{\bf 43}, 971-999, 1995c (Paper III)

{\bf Gr\"un, E. and  Staubach, P.,} Dynamic populations of dust in 
interplanetary space. 
In: {\em Physics, Chemistry and Dynamics of Interplanetary 
Dust}, ASP Conference Series, Vol. 104, (edited by B. A. Gustafson and 
M. S. Hanner), page 3-14, 1996

{\bf Gr\"un, E.,
Baguhl, M., Hamilton, D. P., Riemann, R., Zook, H. A.,
Dermott, S., Fechtig, H., Gustafson, B. A., Hanner, M. S., Hor\'anyi, M.,
Khurana, K. K., Kissel, J., Kivelson, M., Lindblad, B.-A., Linkert, D.,
Linkert, G., Mann, I., McDonnell, J. A. M., Morfill, G. E., Polanskey, C.,
Schwehm, G., and Srama, R.,}
Constraints from Galileo observations on the origin of jovian dust
streams,
{\em Nature}, 381, 395-398, 1996.

{\bf Gr\"un, E., Staubach, P., Baguhl, M., Hamilton, D. P., Zook, H. A., 
Dermott, S., Gustafson, B. A., Fechtig, H., Kissel, J., Linkert, D., 
Linkert, G., Srama, R., Hanner, M. S., Polanskey, C., Horanyi, M., 
Lindblad, B.-A., Mann, I., McDonnell, J. A. M., Morfill, G. E. and
Schwehm, G.,}
South-north and radial traverses through the zodiacal cloud. {\em Icarus},
{\bf 129}, 270-288, 1997
 
{\bf Hamilton, D.P.\ and Burns, J.A.,} Orbital stability zones about
asteroids~II. The destabilizing effects of eccentric orbits and of
solar radiation. {\em Icarus }{\bf 96}, 43-64, 1992

{\bf Hamilton, D.P. and Burns, J.A., } Ejection of dust from Jupiter's
gossamer ring. {\em Nature }{\bf 364}, 695-699, 1993

{\bf Hamilton, D.P., Gr\"un, E. and Baguhl, M.,} Electromagnetic 
escape of dust from the Solar System. 
In: {\em Physics, Chemistry and Dynamics of Interplanetary 
Dust}, ASP Conference Series, Vol. 104, (edited by B. A. Gustafson and 
M. S. Hanner), page 31-34, 1996

{\bf Horanyi, M., Gr\"un, E. and Morfill, G.E., } Mechanism for the 
acceleration and ejection of dust grains from
Jupiter's magnetosphere. {\em Nature }{\bf 363}, 144-146, 1993a

{\bf Horanyi, M., Gr\"un, E. and  Morfill, G.E., } The dusty ballerina
skirt of Jupiter. {\em J.\ Geophys.\ Res. }{\bf 98}, 21245-21251, 1993b

{\bf Kr\"uger, H.,  Gr\"un, E., Hamilton, D. P., Baguhl, M., Dermott, 
S., Fechtig, H., Gustafson, B. A., Hanner, M. S., 
Hor\'anyi, M., Kissel, J., Lindblad, B.-A., Linkert, D., Linkert, G.,
Mann, I., McDonnell, J. A. M., Morfill, G. E., Polanskey, C., 
Riemann, R., Schwehm, G., Srama, R. and Zook, H. A., }
Three years of Galileo dust data: II. 1993 to 1995. 
{\em Planet. Space. Sci.}, 1998, this volume   (Paper~IV) 

{\bf Landgraf, M.,} Modellierung der Dynamik und Interpretation 
der In-Situ-Messung interstellaren Staubs in der lokalen Umgebung 
des Sonnensystems. PhD thesis, University of Heidelberg, 1998

{\bf Landgraf, M. and Gr\"un, E.,} In situ measurements of interstellar 
dust. In: {\em Proceedings of the IAU Colloquium No. 166 on The 
Local Bubble and Beyond}, (edited by D. Breitschwerdt, M.J. Freyberg 
and J. Tr\"umper), Lecture Notes in Physics, Vol. 506, Springer 
Heidelberg, p. 381-384, 1998

{\bf Mann, I., Gr\"un, E., Baguhl, M., Fechtig, H., Hanner, M.S., Kissel, J.,
Lindblad, B.-A., Linkert, D., McDonnell, J.A.M., Morfill, G.E., Polanskey, C.,
Riemann, R., Schwehm, G., Siddique, N. and Zook, H.A., } Measurements with
the Ulysses and Galileo dust detectors close to the ecliptic. In: {\em Proceedings
of the 30th Liege international astrophysical colloquium ``Observations and
physical properties of small solar system bodies'', June 1992},
Univ. de Liege, Institut d'Astrophysique, 13-17, 1992

{\bf Mann, I. and Gr\"un, E., } Dust particles beyond the asteroid
belt -- a study based on recent results of the Ulysses dust
experiment. {\em Planet.\ Space Sci.} {\bf 43}, 827-832, 1995

{\bf Mann, I., Gr\"un, E., Wilk, M.,} The contribution of asteroid dust to the
interplanetary dust cloud: The impact of Ulysses results on the understanding of
dust production in the asteroid belt and the formation of the IRAS dust bands.
{\em Icarus} 120, 399-407, 1996a

{\bf Mann, I., Wilk, M., Gr\"un, E.,} Analysis of Ulysses dust measurements 
within 
the asteroid belt. In: {\em Physics, Chemistry and Dynamics of Interplanetary
Dust}, ASP Conference Series, Vol. 104, (edited by B. A. Gustafson and
M. S. Hanner), page 19-22, 1996b

{\bf Riemann, R.\ and Gr\"un, E.,} Meteor streams, asteroids
and comets near the orbits of Galileo and Ulysses. In: {\em Proceeding
of the workshop on Hypervelocity Impacts in Space}, (edited by J.A.M.\
McDonnell), University of Kent at Canterbury, 120-125, 1992

{\bf Stone, R.G., Bougeret, J.L., Caldwell, J., Canu, P., de
Conchy, Y., Cornilleau-Wehrlin, N., Desch, M.D., Fainberg, J., Goetz, K.,
Goldstein, M.L., Harvey, C.C., Hoang, S., Howard, R., Kaiser, M.L.,
Kellogg, P., Klein, B., Knoll, R., Lecacheux, A., Langyel-Frey, D.,
MacDowall, R.J., Manning, R., Meetre, C.A., Meyer, A., Monge, N.,
Monson, S., Nicol, G., Reiner, M.J., Steinbert, J.L., Torres, E., de
Villedary, C., Wouters, F. and Zarka, P.,}
The unified radio and plasma wave investigation. 
{\em Astron.\ Astrophys.\ Suppl.\ Ser.\ }{\bf 92}, 291-316, 1992

{\bf Svestka, J., Auer, S., Baguhl, M. and Gr\"un, E.,} 
Measurements of dust electric charges by the Ulysses and Galileo 
dust detectors. In: {\em Physics, Chemistry and Dynamics of Interplanetary 
Dust}, ASP Conference Series, Vol. 104, (edited by B. A. Gustafson and 
M. S. Hanner), page 31-34, 1996

{\bf Taylor, A. D., McDonnell, J. A. M. and Gr\"un, E.,} Taurid
complex meteoroids detected near aphelion with Ulysses. 
{\em Adv. Space Res.} 17, No. 12, (12)171-(12)175, 1996

{\bf Wenzel, K.P., Marsden, R.G., Page, D.E. and Smith, E.J.,}
The Ulysses mission. 
{\em Astron.\ Astrophys.\ Suppl.\ Ser. }{\bf 92}, 207-219, 1992

{\bf Witte, M., Banaszkiewicz, H. and Rosenbauer, H.,}
Recent results on the parameters of interstellar helium from the 
Ulysses/GAS experiment,
{\em Space Sci. Rev.} {\bf 78}, No. 1/2, 289-296, 1996

{\bf Zook, H. A., Gr\"un, E., Baguhl, M., Hamilton, D. P., 
Linkert, G., Liou, J.-C., Forsyth, R. and Phillips, J. L.,}
Solar wind magnetic field bending of jovian dust trajectories, 
{\em Science} {\bf 274}, 1501-1503, 1996.

}

\clearpage

\begin{table}[htb]
{\small
\caption{\label{event_table} Ulysses mission and dust detector (GRU) 
configuration, tests and other events. 
Only selected events before 1993 are given. See Sect.~\ref{mission} for 
details.
}
  \begin{tabular*}{15.2cm}{lccl}
   \hline
   \hline \\[-2.0ex]
Yr-day &  
Date & 
Time &  
Event \\[0.7ex]
\hline \\[-2.0ex]
90-279 & 06.10.90 &       & Ulysses launch \\
91-154 & 03.06.91 & 18:13 & GRU heater on: 1,200~mW \\
92-041 & 10.02.92 & 17:00 & GRU nominal configuration: HV=3, EVD=C,I, SSEN=0001 \\
93-045 & 14.02.93 & 06:53 & Ulysses  DNEL \#2 \\
93-045 & 14.02.93 & 22:50 & GRU on, nominal configuration \\
93-120 & 30.04.93 & 02:46 & GRU new program (FN7 data) load \\
93-126 & 06.05.93 & 23:07 & GRU start new program \\
93-196 & 15.07.93 & 02:00 & GRU noise test \\
93-206 & 25.07.93 & 01:00 & GRU HV=4 \\
93-211 & 30.07.93 & 06:03 & GRU HV=3 \\
93-221 & 09.08.93 & 12:17 & Ulysses DNEL \#3 \\
93-223 & 11.08.93 & 22:36 & GRU on \\
93-224 & 12.08.93 & 07:43 & GRU nominal configuration \\
93-231 & 19.08.93 & 01:59 & GRU noise test \\
93-259 & 16.09.93 & 11:59 & GRU noise test \\
93-287 & 14.10.93 & 18:59 & GRU noise test \\
93-315 & 11.11.93 & 19:59 & GRU noise test \\
93-331 & 27.11.93 & 23:41 & Ulysses DNEL \#4 \\
93-332 & 28.11.93 & 15:21 & GRU on \\
93-332 & 28.11.93 & 22:58 & GRU nominal configuration \\
93-343 & 09.12.93 & 18:00 & GRU noise test \\
94-006 & 06.01.94 & 16:00 & GRU noise test \\
94-034 & 03.02.94 & 14:59 & GRU noise test \\
94-062 & 03.03.94 & 07:00 & GRU noise test \\
94-090 & 31.03.94 & 08:45 & GRU noise test \\
94-118 & 28.04.94 & 07:30 & GRU noise test \\
94-146 & 26.05.94 & 05:00 & GRU noise test \\
94-174 & 22.06.94 & 06:00 & GRU noise test \\
94-202 & 21.07.94 & 03:59 & GRU noise test \\
94-230 & 18.08.94 & 02:59 & GRU noise test \\
94-258 & 15.09.94 & 00:00 & GRU noise test \\
94-276 & 03.10.94 &       & Ulysses South polar pass (lat. $-79^{\circ}$, 2.1AU) \\
94-281 & 08.10.94 & 00:30 & GRU heater: 800~mW \\
94-283 & 10.10.94 & 18:45 & Ulysses DNEL \#5 \\
94-284 & 11.10.94 & 23:06 & GRU on \\
94-285 & 12.10.94 & 04:17 & GRU nominal configuration \\
94-286 & 13.10.94 & 19:59 & GRU noise test \\
94-314 & 10.11.94 & 04:59 & GRU noise test \\
94-341 & 08.12.94 & 23:59 & GRU noise test \\
95-005 & 05.01.95 & 00:00 & GRU noise test \\
\hline
\end{tabular*}\\[1.5ex]
}
\end{table}

\begin{table}[t]
{\small
  \begin{tabular*}{15.2cm}{lccl}
\multicolumn{3}{l}{\normalsize{Table~\ref{event_table} continued.}}  &  \\
   \hline
   \hline \\[-2.0ex]
Yr-day & 
Date & 
Time & 
Event \\[0.7ex]
\hline \\[-2.0ex]
95-033 & 02.02.95 & 01:59 & GRU noise test \\
95-036 & 05.02.95 & 01:24 & GRU heater off \\ 
95-071 & 12.03.95 & 12:01 & Ulysses ecliptic plane crossing and perihelion (1.3 AU) \\
95-152 & 01.06.95 & 03:59 & GRU noise test \\
95-163 & 12.06.95 & 03:31 & GRU heater on: 400~mW \\
95-180 & 29.06.95 & 03:59 & GRU noise test \\
95-208 & 27.07.95 & 04:00 & GRU noise test \\
95-216 & 04.08.95 & 00:51 & GRU heater: 800~mW \\
95-231 & 19.08.95 &       & Ulysses North polar pass (lat. $+79^{\circ}$, 2.1AU) \\
95-236 & 24.08.95 & 15:59 & GRU noise test \\
95-264 & 21.09.95 & 18:59 & GRU noise test \\
95-292 & 19.10.95 & 19:59 & GRU noise test \\
95-300 & 27.10.95 & 16:23 & GRU heater: 1,200~mW \\
95-320 & 16.11.95 & 09:59 & GRU noise test \\
95-344 & 10.12.95 & 15:55 & Ulysses DNEL \# 6 \\
95-344 & 10.12.95 & 21:28 & GRU on  \\
95-346 & 12.12.95 & 08:54 & GRU nominal configuration \\
95-348 & 14.12.95 & 15:00 & GRU noise test \\[0.7ex] 
\hline
\end{tabular*}\\[1.5ex]
Abbreviations used: DNEL: Disconnect non-essential loads (i.~e.
all scientific instruments); 
HV: channeltron high voltage step; EVD: event definition,
ion- (I), channeltron- (C), or electron-channel (E); SSEN: detection thresholds,
ICP, CCP, ECP and PCP
}
\end{table}

\clearpage

\pagestyle{empty}


\begin{sidewaystable}
\tiny
\vbox{
\hspace{-3cm}
\begin{minipage}[t]{22cm}
{\bf Table 2.} Overview of dust impacts accumulated with the Ulysses
dust detector between 1 January 1993 and 31 December 1995. Switch-on
of the instrument is indicated by horizontal lines.  The heliocentric
distance R, the lengths of the time interval $\Delta $t (days) from
the previous table entry, and the corresponding numbers of impacts are
given for the 24 accumulators. The accumulators are arranged with
increasing signal amplitude ranges (AR), with four event classes for
each amplitude range (CLN = 0,1,2,3); e.g.~AC31 means counter for AR =
1 and CLN = 3.  The $\Delta $t in the first line (93-001) is the time
interval counted from the last entry in Table~3 in Paper~III. The
totals of counted impacts$^{\ast}$, of impacts with complete data, and
of all events (noise plus impact events) for the entire period are
given as well.
\end{minipage}
}
\bigskip
\hspace{-3cm}
\begin{tabular}{|r|r|r|r|cccc|cccc|cccc|cccc|cccc|cccc|}
\hline
&&&&& &&&&& &&&&& &&&&& &&&&& &&\\
\mc{1}{|c|}{Date}&
\mc{1}{|c|}{Time}&
\mc{1}{|c|}{R}&
\mc{1}{|c|}{$\Delta $t}&
{\scriptsize AC}&{\scriptsize AC}&{\scriptsize AC}&{\scriptsize AC}&
{\scriptsize AC}&{\scriptsize AC}&{\scriptsize AC}&{\scriptsize AC}&
{\scriptsize AC}&{\scriptsize AC}&{\scriptsize AC}&{\scriptsize AC}&
{\scriptsize AC}&{\scriptsize AC}&{\scriptsize AC}&{\scriptsize AC}&
{\scriptsize AC}&{\scriptsize AC}&{\scriptsize AC}&{\scriptsize AC}&
{\scriptsize AC}&{\scriptsize AC}&{\scriptsize AC}&{\scriptsize AC}\\
&&[AU]&\mc{1}{c|}{[d]}&
\,\,01$^\ast$&\,\,11$^\ast$&21&31&
\,\,02$^\ast$&12&22&32&
03&13&23&33&
04&14&24&34&
05&15&25&35&
06&16&26&36\\
&&&&& &&&&& &&&&& &&&&& &&&&& &&\\
\hline
&&&&& &&&&& &&&&& &&&&& &&&&& &&\\
93-001& 02:08&      5.065& 2.1&-&-&-&-&-&-&-&-&-&-&-&-&-&-&-&-&-&-&-&-&-&-&-&-\\
93-008& 09:24&      5.049& 7.3&-& 1&-&-&-&-&-& 2&-&-&-& 2&-&-&-&-&-&-&-&-&-&-&-&-\\
93-016& 02:51&      5.028& 7.7& 1&-&-&-& 1&-&-& 2&-&-&-&-&-&-&-&-&-&-&-&-&-&-&-&-\\
93-023& 03:00&      5.011& 7.0& 1&-&-&-&-& 1&-&-&-&-&-&-&-&-&-&-&-&-&-&-&-&-&-&-\\
93-030& 07:15&      5.001& 7.2&-&-&-&-&-&-&-& 1&-&-&-& 1&-&-&-&-&-&-&-&-&-&-&-&-\\
 &&&&& &&&&& &&&&& &&&&& &&&&& &&\\
93-044& 18:50&      4.967& 14.5& 1& 1&-& 1&-& 1&-& 1&-&-&-&-&-&-&-&-&-&-&-&-&-&-&-&-\\
93-045& 14:53&      4.952& 0.8&\mc{4}{|c|}{---------------------}&\mc{4}{|c|}{---------------------}&\mc{4}{|c|}{---------------------}&\mc{4}{|c|}{---------------------}&\mc{4}{|c|}{---------------------}&\mc{4}{|c|}{---------------------}\\
93-053& 02:13&      4.931& 7.5&-&-&-&-& 1&-&-& 2&-& 1&-& 2&-&-&-&-&-&-&-&-&-&-&-&-\\
93-060& 06:07&      4.929& 7.2&-&-&-&-&-&-&-&-&-&-&-&-&-&-&-&-&-&-&-&-&-&-&-&-\\
93-068& 05:29&      4.888& 8.0& 2&-&-&-&-&-&-&-&-&-&-&-&-&-&-&-&-&-&-&-&-&-&-&-\\
 &&&&& &&&&& &&&&& &&&&& &&&&& &&\\
93-076& 01:48&      4.889& 7.8& 1&-&-&-& 1&-&-& 2&-&-&-& 1&-&-&-&-&-&-&-&-&-&-&-&-\\
93-083& 03:48&      4.870& 7.1& 1&-&-&-&-&-&-& 2&-&-&-& 1&-&-&-&-&-&-&-&-&-&-&-&-\\
93-090& 04:04&      4.852& 7.0&-& 1&-&-& 1&-&-& 3&-&-&-& 1&-&-&-&-&-&-&-&-&-&-&-&-\\
93-098& 02:25&      4.796& 7.9& 1&-&-&-&-& 2&-& 1&-&-&-&-&-&-&-&-&-&-&-&-&-&-&-&-\\
93-105& 04:50&      4.810& 7.1& 1& 1&-&-&-& 1&-&-&-&-&-&-&-&-&-&-&-&-&-&-&-&-&-&-\\
 &&&&& &&&&& &&&&& &&&&& &&&&& &&\\
93-113& 00:08&      4.788& 7.8&-& 1&-&-&-&-&-& 1&-&-&-&-&-&-&-&-&-&-&-&-&-&-&-&-\\
93-120& 00:37&      4.768& 7.0& 1&-&-&-&-&-&-& 2&-&-&-&-&-&-&-& 1&-&-&-&-&-&-&-&-\\
93-127& 01:10&      4.747& 7.0& 1&-&-&-&-&-& 1&-&-&-&-&-&-&-&-&-&-&-&-&-&-&-&-&-\\
93-134& 04:48&      4.726& 7.2&-&-&-& 1&-&-&-&-&-&-&-&-&-&-&-&-&-&-&-&-&-&-&-&-\\
93-142& 00:32&      4.702& 7.8&-&-&-&-&-&-&-&-&-&-&-&-&-&-&-&-&-&-&-&-&-&-&-&-\\
 &&&&& &&&&& &&&&& &&&&& &&&&& &&\\
93-149& 01:25&      4.680& 7.0&-&-&-&-&-& 1&-&-&-&-&-&-&-&-&-&-&-&-&-&-&-&-&-&-\\
93-156& 02:37&      4.597& 7.1& 1&-&-&-&-&-&-&-&-&-&-& 1&-&-&-&-&-&-&-&-&-&-&-&-\\
93-164& 01:55&      4.633& 8.0& 1& 2&-&-&-&-&-&-&-&-&-& 1&-&-&-&-&-&-&-&-&-&-&-&-\\
93-171& 02:42&      4.540& 7.0&-&-&-&-&-&-&-&-&-&-&-&-&-&-&-&-&-&-&-&-&-&-&-&-\\
93-179& 01:33&      4.583& 8.0&-&-&-&-&-&-&-& 1&-&-&-& 1&-&-&-&-&-&-&-&-&-&-&-&-\\
 &&&&& &&&&& &&&&& &&&&& &&&&& &&\\
93-187& 11:29&      4.476& 8.4&-&-&-&-&-&-&-&-&-&-&-&-&-&-&-&-&-&-&-&-&-&-&-&-\\
93-195& 01:06&      4.529& 7.6&-&-&-&-&-&-&-& 1&-&-&-&-&-&-&-&-&-&-&-&-&-&-&-&-\\
93-202& 01:49&      4.505& 7.0&-&-&-&-&-& 1&-&-&-&-&-&-&-&-&-&-&-&-&-&-&-&-&-&-\\
93-210& 00:09&      4.476& 7.9&-&-&-&-&-&-&-&-&-&-&-& 1&-&-&-&-&-&-&-&-&-&-&-&-\\
93-217& 00:11&      4.451& 7.0&-&-&-&-&-& 1&-&-&-&-&-& 1&-&-&-&-&-&-&-&-&-&-&-&-\\
 &&&&& &&&&& &&&&& &&&&& &&&&& &&\\
93-224& 06:14&      4.424& 7.3& 1&-&-&-&-&-&-&-&-&-&-& 1&-&-&-&-&-&-&-&-&-&-&-&-\\
93-225& 06:29&      4.421& 1.0&\mc{4}{|c|}{---------------------}&\mc{4}{|c|}{---------------------}&\mc{4}{|c|}{---------------------}&\mc{4}{|c|}{---------------------}&\mc{4}{|c|}{---------------------}&\mc{4}{|c|}{---------------------}\\
93-233& 01:23&      4.392& 7.8& 1&-&-&-&-&-&-& 1&-&-&-&-&-&-&-&-&-&-&-&-&-&-&-&-\\
93-241& 03:55&      4.361& 8.1& 1&-&-&-&-&-& 1& 1&-&-&-& 2&-&-&-&-&-&-&-&-&-&-&-&-\\
93-249& 02:41&      4.210& 7.9& 1&-&-& 1&-&-&-&-&-&-&-&-&-&-&-&-&-&-&-&-&-&-&-&-\\
 &&&&& &&&&& &&&&& &&&&& &&&&& &&\\
\hline
\end{tabular}
\end{sidewaystable}
\clearpage
\begin{sidewaystable}
\tiny
\hspace{-3cm}
\begin{tabular}{|r|r|r|r|cccc|cccc|cccc|cccc|cccc|cccc|}
\hline
&&&&& &&&&& &&&&& &&&&& &&&&& &&\\
\mc{1}{|c|}{Date}&
\mc{1}{|c|}{Time}&
\mc{1}{|c|}{R}&
\mc{1}{|c|}{$\Delta $t}&
{\scriptsize AC}&{\scriptsize AC}&{\scriptsize AC}&{\scriptsize AC}&
{\scriptsize AC}&{\scriptsize AC}&{\scriptsize AC}&{\scriptsize AC}&
{\scriptsize AC}&{\scriptsize AC}&{\scriptsize AC}&{\scriptsize AC}&
{\scriptsize AC}&{\scriptsize AC}&{\scriptsize AC}&{\scriptsize AC}&
{\scriptsize AC}&{\scriptsize AC}&{\scriptsize AC}&{\scriptsize AC}&
{\scriptsize AC}&{\scriptsize AC}&{\scriptsize AC}&{\scriptsize AC}\\
&&[AU]&\mc{1}{c|}{[d]}&
\,\,01$^\ast$&\,\,11$^\ast$&21&31&
\,\,02$^\ast$&12&22&32&
03&13&23&33&
04&14&24&34&
05&15&25&35&
06&16&26&36\\
&&&&& &&&&& &&&&& &&&&& &&&&& &&\\
\hline
&&&&& &&&&& &&&&& &&&&& &&&&& &&\\
93-257& 01:23&      4.299& 7.9&-&-&-&-& 1&-&-&-&-&-&-&-&-&-&-&-&-&-&-&-&-&-&-&-\\
93-265& 10:20&      4.266& 8.4&-&-&-&-&-& 1&-& 1&-&-&-&-&-&-&-&-&-&-&-&-&-&-&-&-\\
93-273& 01:04&      4.235& 7.6& 2&-&-&-&-&-&-& 2&-&-&-&-&-&-&-&-&-&-&-&-&-&-&-&-\\
93-281& 01:15&      4.202& 8.0&-&-&-&-&-& 1& 1&-&-&-&-& 1&-&-&-&-&-&-&-&-&-&-&-&-\\
93-288& 02:55&      4.021& 7.1& 1&-&-&-& 1&-&-&-&-&-&-&-&-&-&-&-&-&-&-&-&-&-&-&-\\
 &&&&& &&&&& &&&&& &&&&& &&&&& &&\\
93-296& 06:11&      4.138& 8.1&-& 1&-&-&-& 1&-&-&-&-&-&-&-&-&-&-&-&-&-& 1&-&-&-&-\\
93-304& 00:43&      4.105& 7.8&-&-&-&-&-&-&-& 1&-&-&-& 2&-&-&-& 1&-&-&-&-&-&-&-&-\\
93-312& 02:43&      3.898& 8.1& 1&-&-&-&-&-&-&-&-&-&-& 1&-&-&-&-&-&-&-&-&-&-&-&-\\
93-320& 03:49&      4.034& 8.0& 2&-&-&-&-&-&-& 2&-&-&-&-&-&-&-&-&-&-&-&-&-&-&-&-\\
93-331& 17:21&      3.792& 11.6& 4&-&-&-&-& 1&-& 1&-&-&-& 1&-& 1&-&-&-&-&-&-&-&-&-&-\\
 &&&&& &&&&& &&&&& &&&&& &&&&& &&\\
93-332& 17:22&      3.786& 1.0&\mc{4}{|c|}{---------------------}&\mc{4}{|c|}{---------------------}&\mc{4}{|c|}{---------------------}&\mc{4}{|c|}{---------------------}&\mc{4}{|c|}{---------------------}&\mc{4}{|c|}{---------------------}\\
93-340& 01:29&      3.943& 7.3&-&-&-&-&-&-&-&-&-&-&-&-&-&-&-&-&-&-&-&-&-&-&-&-\\
93-348& 00:46&      3.906& 8.0& 1&-&-&-& 1&-&-&-&-&-&-& 2&-&-&-& 1&-&-&-&-&-&-&-&-\\
93-356& 00:16&      3.868& 8.0&-&-&-&-&-&-&-&-&-&-&-& 1&-&-&-&-&-&-&-&-&-&-&-&-\\
93-363& 00:49&      3.834& 7.0&-&-&-&-&-&-&-&-&-&-&-&-&-&-&-&-&-&-&-&-&-&-&-&-\\
 &&&&& &&&&& &&&&& &&&&& &&&&& &&\\
94-006& 01:02&      3.795& 8.0& 1& 1&-&-&-&-&-&-&-&-&-& 2&-&-&-&-&-&-&-&-&-&-&-&-\\
94-014& 01:59&      3.755& 8.0&-&-&-&-&-&-&-& 1&-&-&-& 1&-&-&-&-&-&-&-&-&-&-&-&-\\
94-022& 00:58&      3.715& 8.0& 3&-&-& 1&-&-&-&-&-&-&-&-&-&-&-& 1&-&-&-&-&-&-&-&-\\
94-030& 00:15&      3.674& 8.0& 2&-&-&-&-&-&-& 1&-&-&-&-&-&-&-&-&-&-&-&-&-&-&-&-\\
94-037& 02:40&      3.614& 7.1&-&-&-&-&-&-&-&-&-&-&-& 3&-&-&-&-&-&-&-&-&-&-&-&-\\
 &&&&& &&&&& &&&&& &&&&& &&&&& &&\\
94-045& 00:18&      3.596& 7.9& 1&-&-&-&-& 1&-&-&-&-&-& 1&-&-&-&-&-&-&-&-&-&-&-&-\\
94-052& 00:41&      3.559& 7.0&-&-&-&-&-&-&-&-&-&-&-&-&-&-&-&-&-&-&-&-&-&-&-&-\\
94-060& 01:07&      3.516& 8.0& 1&-&-&-&-&-&-&-&-&-&-&-&-&-&-&-&-&-&-&-&-&-&-&-\\
94-068& 00:22&      3.473& 8.0&-&-&-&-&-&-&-&-&-&-&-& 1&-&-&-&-&-&-&-&-&-&-&-&-\\
94-075& 00:42&      3.434& 7.0& 1&-&-&-&-&-&-&-&-&-&-& 1&-&-&-&-&-&-&-&-&-&-&-&-\\
 &&&&& &&&&& &&&&& &&&&& &&&&& &&\\
94-083& 00:55&      3.389& 8.0& 1&-&-&-&-&-&-&-&-&-&-&-&-&-&-&-&-&-&-&-&-&-&-&-\\
94-090& 01:14&      3.350& 7.0& 4&-&-& 1&-& 1&-&-&-& 1&-&-&-&-&-&-&-&-&-&-&-&-&-&-\\
94-098& 01:19&      3.304& 8.0& 1&-&-&-&-&-&-&-&-& 1&-&-&-&-&-&-&-&-&-&-&-&-&-&-\\
94-106& 00:45&      3.257& 8.0& 2&-&-&-& 1&-&-&-&-&-&-& 1&-&-&-&-&-&-&-&-&-&-&-&-\\
94-113& 00:46&      3.216& 7.0& 2&-&-&-&-&-&-&-&-&-& 1&-&-&-&-&-&-&-&-&-&-&-&-&-\\
 &&&&& &&&&& &&&&& &&&&& &&&&& &&\\
94-124& 00:05&      3.150& 11.0& 1&-&-&-&-&-&-&-&-& 1&-&-&-&-&-&-&-&-&-& 1&-&-&-&-\\
94-131& 00:15&      3.108& 7.0& 1&-&-&-&-&-&-&-&-&-&-&-&-&-&-&-&-&-&-&-&-&-&-&-\\
94-138& 01:26&      3.065& 7.0& 1&-&-&-&-&-&-&-&-&-&-& 1&-&-&-&-&-&-&-&-&-&-&-&-\\
94-146& 00:26&      3.016& 8.0&-&-&-&-& 2& 1&-&-&-&-&-&-&-&-&-&-&-&-&-&-&-&-&-&-\\
94-154& 00:44&      2.966& 8.0&-&-&-&-&-&-&-&-&-&-&-& 2&-&-&-&-&-&-&-&-&-&-&-&-\\
 &&&&& &&&&& &&&&& &&&&& &&&&& &&\\
94-162& 00:27&      2.915& 8.0& 2&-&-&-&-&-& 1&-&-&-&-&-&-&-&-&-&-&-&-&-&-&-&-&-\\
94-169& 23:56&      2.728& 8.0& 1&-&-&-& 1&-&-&-&-&-&-&-&-&-&-&-&-&-&-&-&-&-&-&-\\
94-177& 00:04&      2.818& 7.0& 1&-&-&-&-& 1&-&-&-&-&-& 1&-&-&-&-&-&-&-&-&-&-&-&-\\
94-184& 00:12&      2.773& 7.0& 4&-&-& 1& 1&-&-& 1&-&-& 1&-&-&-&-&-&-&-&-&-&-&-&-&-\\
94-192& 00:11&      2.720& 8.0& 3& 1&-&-&-& 1&-&-&-&-&-&-&-&-&-&-&-&-&-&-&-&-&-&-\\
 &&&&& &&&&& &&&&& &&&&& &&&&& &&\\
94-199& 02:39&      2.506& 7.1& 1&-&-&-& 1& 2&-&-&-&-&-& 1&-&-&-&-&-&-&-&-&-&-&-&-\\
94-207& 00:07&      2.619& 7.9& 3&-&-&-&-&-&-&-&-&-&-& 1&-&-&-&-&-&-&-&-&-&-&-&-\\
94-214& 00:09&      2.572& 7.0&-&-&-& 1&-& 1&-&-&-&-&-&-&-&-&-&-&-&-&-&-&-&-&-&-\\
94-221& 00:43&      2.524& 7.0&-&-&-&-&-& 1&-&-&-&-&-&-&-&-&-&-&-&-&-&-&-&-&-&-\\
94-229& 00:55&      2.469& 8.0& 3& 1&-&-&-&-&-&-&-&-&-&-&-&-&-&-&-&-&-& 1&-&-&-&-\\
 &&&&& &&&&& &&&&& &&&&& &&&&& &&\\
\hline
\end{tabular}
\end{sidewaystable}
\clearpage
\begin{sidewaystable}
\tiny
\hspace{-3cm}
\begin{tabular}{|r|r|r|r|cccc|cccc|cccc|cccc|cccc|cccc|}
\hline
&&&&& &&&&& &&&&& &&&&& &&&&& &&\\
\mc{1}{|c|}{Date}&
\mc{1}{|c|}{Time}&
\mc{1}{|c|}{R}&
\mc{1}{|c|}{$\Delta $t}&
{\scriptsize AC}&{\scriptsize AC}&{\scriptsize AC}&{\scriptsize AC}&
{\scriptsize AC}&{\scriptsize AC}&{\scriptsize AC}&{\scriptsize AC}&
{\scriptsize AC}&{\scriptsize AC}&{\scriptsize AC}&{\scriptsize AC}&
{\scriptsize AC}&{\scriptsize AC}&{\scriptsize AC}&{\scriptsize AC}&
{\scriptsize AC}&{\scriptsize AC}&{\scriptsize AC}&{\scriptsize AC}&
{\scriptsize AC}&{\scriptsize AC}&{\scriptsize AC}&{\scriptsize AC}\\
&&[AU]&\mc{1}{c|}{[d]}&
\,\,01$^\ast$&\,\,11$^\ast$&21&31&
\,\,02$^\ast$&12&22&32&
03&13&23&33&
04&14&24&34&
05&15&25&35&
06&16&26&36\\
&&&&& &&&&& &&&&& &&&&& &&&&& &&\\
\hline
&&&&& &&&&& &&&&& &&&&& &&&&& &&\\
94-237& 00:24&      2.413& 8.0& 2&-&-& 1& 1& 1&-& 1&-&-&-&-&-&-&-&-&-&-&-&-&-&-&-&-\\
94-245& 00:24&      2.357& 8.0& 3&-&-& 1&-&-&-&-&-&-&-&-&-&-&-&-&-&-&-&-&-&-&-&-\\
94-253& 00:24&      2.301& 8.0&-&-&-&-&-& 1&-&-&-&-&-& 2&-&-&-&-&-&-&-&-&-&-&-&-\\
94-261& 00:02&      2.245& 8.0& 1&-&-& 1&-&-&-&-&-&-&-& 1&-&-&-&-&-&-&-&-&-&-&-&-\\
94-268& 00:30&      2.195& 7.0& 1&-&-&-&-&-&-& 1&-&-&-&-&-&-&-&-&-&-&-&-&-&-&-&-\\
 &&&&& &&&&& &&&&& &&&&& &&&&& &&\\
94-276& 00:00&      1.895& 8.0& 1&-&-&-&-&-&-& 1&-& 1&-&-&-&-&-&-&-&-&-&-&-&-&-& 1\\
94-283& 15:14&      2.084& 7.6& 4&-&-&-& 1&-&-&-&-&-&-& 1&-&-&-&-&-&-&-&-&-&-&-&-\\
94-285& 23:58&      1.816& 2.4&\mc{4}{|c|}{---------------------}&\mc{4}{|c|}{---------------------}&\mc{4}{|c|}{---------------------}&\mc{4}{|c|}{---------------------}&\mc{4}{|c|}{---------------------}&\mc{4}{|c|}{---------------------}\\
94-293& 00:02&      2.018& 7.0& 1& 1&-& 1& 1&-&-&-&-&-&-& 1&-&-&-& 1&-&-&-&-&-&-&-&-\\
94-301& 00:16&      1.962& 8.0& 1&-&-&-&-&-&-& 2&-&-&-& 1&-&-&-&-&-&-&-&-&-&-&-&-\\
 &&&&& &&&&& &&&&& &&&&& &&&&& &&\\
94-309& 00:01&      1.906& 8.0& 5&-&-&-&-& 2&-& 1&-&-& 1&-&-&-&-&-&-&-&-&-&-&-&-&-\\
94-317& 00:02&      1.851& 8.0& 5&-&-&-&-& 1&-& 1&-&-&-&-&-&-&-&-&-&-&-&-&-&-&-&-\\
94-324& 00:36&      1.804& 7.0& 2&-&-& 1&-&-&-&-&-&-&-&-&-&-&-& 1&-&-&-&-&-&-&-&-\\
94-331& 06:16&      1.755& 7.2& 1&-&-&-& 1& 1&-&-&-&-&-&-&-&-&-& 1&-&-&-&-&-&-&-&-\\
94-339& 01:47&      1.705& 7.8& 1&-&-&-&-& 1&-& 1&-&-&-& 1&-&-& 1&-&-&-&-&-&-&-&-&-\\
 &&&&& &&&&& &&&&& &&&&& &&&&& &&\\
94-347& 00:10&      1.655& 7.9& 3&-&-&-& 1&-&-&-&-& 1&-& 1&-&-&-&-&-&-&-&-&-&-&-&-\\
94-354& 00:44&      1.613& 7.0& 1&-&-&-& 1&-&-&-& 1&-&-&-&-&-&-&-&-&-&-&-&-& 1&-&-\\
94-362& 00:00&      1.314& 8.0& 1&-&-&-& 1&-&-&-&-&-&-&-&-&-& 1& 1&-&-&-&-&-&-&-&-\\
95-004& 01:17&      1.529& 7.1&-&-&-&-& 1&-&-&-&-&-&-&-&-&-&-&-&-&-&-&-&-&-&-&-\\
95-012& 00:02&      1.488& 7.9& 1&-&-&-&-& 1&-&-&-&-& 1&-&-&-&-& 1&-&-&-&-&-&-&-&-\\
 &&&&& &&&&& &&&&& &&&&& &&&&& &&\\
95-019& 23:57&      1.442& 8.0& 3&-&-&-& 1&-&-&-&-&-&-&-&-&-& 1& 1&-&-&-&-&-&-&-&-\\
95-027& 00:04&      1.423& 7.0& 5&-&-&-&-&-&-&-&-&-&-& 1&-&-&-&-&-&-&-&-&-&-&-& 1\\
95-034& 23:59&      1.382& 8.0& 5&-&-&-&-&-& 1&-&-&-&-&-&-&-&-&-&-&-&-&-&-&-&-&-\\
95-042& 01:28&      1.375& 7.1& 3&-&-&-&-&-&-& 1&-&-&-&-&-&-&-&-&-&-&-&-&-&-&-&-\\
95-050& 01:03&      1.357& 8.0& 2&-&-&-&-&-&-&-&-&-&-&-&-&-&-&-&-& 1&-&-&-&-&-&-\\
 &&&&& &&&&& &&&&& &&&&& &&&&& &&\\
95-058& 00:57&      1.346& 8.0& 5&-&-&-& 1&-&-& 1& 1&-&-& 1&-& 1&-&-&-&-&-&-&-&-&-&-\\
95-066& 00:01&      1.340& 8.0& 7& 1&-& 1&-&-&-&-&-&-&-& 1&-& 2&-&-&-& 1&-&-&-&-&-&-\\
95-073& 00:41&      1.340& 7.0& 11&-&-&-& 2&-&-&-&-& 1&-& 1&-&-&-& 1&-&-& 1& 1&-&-&-&-\\
95-080& 01:36&      1.346& 7.0& 7&-&-&-&-&-&-&-&-&-& 1&-&-&-&-&-&-&-&-&-&-& 1&-&-\\
95-088& 00:29&      1.358& 8.0& 6&-&-&-& 3&-&-&-&-&-&-& 1&-&-&-&-&-&-&-&-&-&-&-& 2\\
 &&&&& &&&&& &&&&& &&&&& &&&&& &&\\
95-095& 00:30&      1.373& 7.0& 9&-& 1&-&-&-&-&-&-&-&-& 1&-&-&-&-&-&-& 1& 1&-&-&-&-\\
95-103& 00:08&      1.395& 8.0& 8&-&-&-&-&-&-&-&-&-&-& 2&-&-&-&-&-&-&-&-&-& 1&-&-\\
95-111& 00:00&      1.476& 8.0& 4&-&-&-&-& 1&-&-&-&-&-&-&-&-&-&-&-&-&-&-&-&-&-&-\\
95-119& 00:15&      1.456& 8.0& 3& 1&-&-&-&-&-&-& 1&-&-&-&-&-&-&-&-&-&-&-&-&-&-&-\\
95-127& 00:30&      1.493& 8.0&-&-&-& 1&-&-&-&-&-&-&-&-&-&-&-&-&-&-&-&-&-&-&-&-\\
 &&&&& &&&&& &&&&& &&&&& &&&&& &&\\
95-135& 00:11&      1.534& 8.0& 2& 1&-& 1&-&-&-&-&-&-&-&-&-&-&-& 1&-&-&-&-&-&-&-&-\\
95-142& 00:35&      1.572& 7.0& 5&-&-&-&-&-&-&-&-&-&-&-&-&-&-&-&-&-&-&-&-&-&-&-\\
95-150& 00:51&      1.619& 8.0& 1&-&-&-& 1&-&-&-&-&-&-&-&-&-&-&-&-&-&-&-&-&-&-&-\\
95-158& 01:05&      1.667& 8.0& 2&-&-&-&-&-&-&-&-&-&-& 1&-&-&-&-&-&-&-&-&-&-&-&-\\
95-166& 01:25&      1.718& 8.0& 1&-&-&-&-&-&-& 1&-&-&-&-&-&-&-&-&-&-&-&-&-&-&-&-\\
 &&&&& &&&&& &&&&& &&&&& &&&&& &&\\
95-174& 01:40&      1.770& 8.0& 2&-&-&-&-&-&-&-&-&-&-&-&-&-&-&-&-&-&-&-&-&-&-&-\\
95-182& 01:15&      1.824& 8.0& 2&-&-&-&-&-&-&-&-&-&-&-&-&-&-&-&-&-&-&-&-&-&-&-\\
95-189& 01:43&      1.872& 7.0&-&-&-& 2& 1&-&-&-&-&-&-&-&-&-&-&-&-&-&-&-&-&-&-&-\\
95-197& 01:05&      1.927& 8.0&-&-&-&-&-&-&-& 1&-&-&-&-&-&-&-&-&-&-&-&-&-&-&-&-\\
95-204& 01:51&      1.976& 7.0& 3&-&-&-&-&-&-&-&-&-&-& 1&-&-&-&-&-&-&-&-&-&-&-&-\\
 &&&&& &&&&& &&&&& &&&&& &&&&& &&\\
\hline
\end{tabular}
\end{sidewaystable}
\clearpage
\begin{sidewaystable}
\tiny
\hspace{-3cm}
\begin{tabular}{|r|r|r|r|cccc|cccc|cccc|cccc|cccc|cccc|}
\hline
&&&&& &&&&& &&&&& &&&&& &&&&& &&\\
\mc{1}{|c|}{Date}&
\mc{1}{|c|}{Time}&
\mc{1}{|c|}{R}&
\mc{1}{|c|}{$\Delta $t}&
{\scriptsize AC}&{\scriptsize AC}&{\scriptsize AC}&{\scriptsize AC}&
{\scriptsize AC}&{\scriptsize AC}&{\scriptsize AC}&{\scriptsize AC}&
{\scriptsize AC}&{\scriptsize AC}&{\scriptsize AC}&{\scriptsize AC}&
{\scriptsize AC}&{\scriptsize AC}&{\scriptsize AC}&{\scriptsize AC}&
{\scriptsize AC}&{\scriptsize AC}&{\scriptsize AC}&{\scriptsize AC}&
{\scriptsize AC}&{\scriptsize AC}&{\scriptsize AC}&{\scriptsize AC}\\
&&[AU]&\mc{1}{c|}{[d]}&
\,\,01$^\ast$&\,\,11$^\ast$&21&31&
\,\,02$^\ast$&12&22&32&
03&13&23&33&
04&14&24&34&
05&15&25&35&
06&16&26&36\\
&&&&& &&&&& &&&&& &&&&& &&&&& &&\\
\hline
&&&&& &&&&& &&&&& &&&&& &&&&& &&\\
95-212& 00:07&      2.032& 7.9& 2& 1&-&-& 1&-&-&-&-& 1&-&-&-&-&-&-&-&-&-&-&-&-&-&-\\
95-219& 00:07&      2.082& 7.0& 1&-&-&-& 3&-&-& 1&-&-&-&-&-&-&-&-&-&-&-&-&-&-&-&-\\
95-226& 01:16&      2.131& 7.0& 2&-&-&-& 1&-&-&-&-&-&-&-&-&-&-&-&-&-&-&-&-&-&-&-\\
95-233& 01:42&      2.181& 7.0& 4&-&-&-& 1&-&-&-&-&-&-&-&-&-&-&-&-&-&-&-&-&-&-&-\\
95-240& 23:59&      2.450& 7.9& 3&-&-&-&-&-&-&-&-&-&-&-&-&-&-&-&-&-&-&-&-&-&-&-\\
 &&&&& &&&&& &&&&& &&&&& &&&&& &&\\
95-248& 00:49&      2.287& 7.0& 2& 1&-&-&-&-&-&-&-&-&-&-&-&-&-&-&-& 1&-&-&-&-&-&-\\
95-255& 01:18&      2.336& 7.0& 1&-&-&-&-&-&-&-&-&-&-&-&-&-&-&-&-&-&-&-&-&-&-&-\\
95-262& 01:50&      2.385& 7.0& 1&-&-&-&-&-&-&-&-&-& 1& 1&-&-&-&-&-&-&-&-&-&-&-&-\\
95-270& 00:19&      2.441& 7.9& 3&-&-&-&-&-&-& 1&-& 1&-&-&-&-&-&-&-&-&-&-&-&-&-&-\\
95-277& 00:19&      2.489& 7.0& 1&-&-&-&-&-&-&-&-&-&-&-&-&-&-&-&-&-&-&-&-&-&-&-\\
 &&&&& &&&&& &&&&& &&&&& &&&&& &&\\
95-284& 01:14&      2.538& 7.0& 2&-&-&-&-&-&-&-&-&-&-&-&-&-&-&-&-&-&-&-&-&-&-&-\\
95-292& 00:08&      2.592& 8.0& 1&-&-&-&-&-&-&-&-&-&-&-&-&-&-&-&-&-&-&-&-&-&-&-\\
95-299& 00:27&      2.640& 7.0& 2&-&-&-& 1&-&-&-&-&-&-& 2&-&-&-&-&-&-&-&-&-&-&-&-\\
95-306& 01:38&      2.687& 7.0& 3&-&-&-&-&-&-&-&-&-&-&-&-&-&-&-&-&-&-&-&-&-&-&-\\
95-313& 01:38&      2.733& 7.0& 1&-&-&-&-&-&-&-&-&-&-&-&-&-&-&-&-&-&-&-&-&-&-&-\\
 &&&&& &&&&& &&&&& &&&&& &&&&& &&\\
95-320& 03:37&      2.780& 7.1& 3&-&-&-&-&-&-&-&-&-&-& 1&-&-&-&-&-&-&-&-&-&-&-&-\\
95-328& 01:21&      2.831& 7.9& 1&-&-&-& 2&-&-&-&-&-&-&-&-&-&-&-&-&-&-&-&-&-&-&-\\
95-335& 02:38&      3.146& 7.1&-&-&-&-&-&-&-&-&-&-&-&-&-&-&-&-&-&-&-&-&-&-&-&-\\
95-344& 13:11&      2.937& 9.4& 1&-&-& 1&-&-&-&-&-&-&-&-&-&-&-&-&-&-&-&-&-&-&-&-\\
95-345& 00:05&      2.941& 0.5&\mc{4}{|c|}{---------------------}&\mc{4}{|c|}{---------------------}&\mc{4}{|c|}{---------------------}&\mc{4}{|c|}{---------------------}&\mc{4}{|c|}{---------------------}&\mc{4}{|c|}{---------------------}\\
 &&&&& &&&&& &&&&& &&&&& &&&&& &&\\
95-352& 01:00&      2.985& 7.0& 1&-&-&-&-&-&-&-&-&-&-& 1&-&-&-&-&-&-&-&-&-&-&-&-\\
95-360& 01:15&      3.035& 8.0& 1&-&-&-& 1&-&-&-&-&-&-&-&-&-&-&-&-&-&-&-&-&-&-&-\\
\hline 
\mc{4}{|l|}{}&& &&&&& &&&&& &&&&& &&&&& &&\\
\mc{4}{|l|}{Impacts (counted)}&\, 235$^{\ast}$ &\, 17$^{\ast}$ & 1& 18&\, 40$^{\ast}$ & 30& 5& 47& 3& 9& 6& 63& 0& 4& 3& 12& 0& 3& 2& 5& 0& 3& 0& 4\\[1.5ex]
\mc{4}{|l|}{Impacts (complete data)}& 235& 17& 1& 18& 40& 30& 5& 47& 2& 9& 6& 63& 0& 4& 3& 12& 0& 3& 2& 5& 0& 3& 0& 4\\[1.5ex]
\mc{4}{|l|}{All events(complete data)}& 68176& 3676& 1& 18& 738& 30& 5& 47& 3& 9& 6& 63& 1& 4& 3& 12& 0& 3& 2& 5& 0& 3& 0& 4\\[1.5ex]
\hline
\end{tabular}\\
\vbox{
\vspace{0.5cm}
\hspace{-3cm}\begin{minipage}[t]{22cm}
$\ast$: {\footnotesize Entries for AC01, AC11 and AC02 are the number of
impacts with complete data. Due to the noise contamination of these
three categories the number of impacts cannot be determined from the
accumulators. The method to separate dust impacts from noise events in
these three categories has been given by Baguhl et al. (1993)}
\end{minipage}}

\end{sidewaystable}

\clearpage
\tabcolsep0.07cm 
\renewcommand{\arraystretch}{0.1}

\begin{sidewaystable}
\tiny
\hspace{-4cm}
\begin{minipage}[t]{22cm}
{\bf Table 3.}
 Raw data: No., impact time, CLN, AR, SEC, IA, EA, CA, IT, ET, 
 EIT, EIC, ICC, PA, PET, EVD, ICP, ECP, CCP, PCP, HV and 
 evaluated data: R, LON, LAT, $D_{\rm Jup}$, rotation angle (ROT), instr. pointing 
 ($\rm S_{LON}$, $\rm S_{LAT}$), speed (V), 
 speed error factor (VEF), mass (M) and mass error factor (MEF).
\end{minipage}

\bigskip

\hspace{-4.0cm}
\begin{tabular*}{26cm} {@{\extracolsep{\fill}}                                  
rrrrrrrrrrrrrrrcrrrrcrrrrrrrrrrrrrrrrrrrrr}                                     
 \mc{1}{c}{No.}&               \mc{1}{c}{IMP. DATE}&         \mc{1}{c}{C}&                 \mc{1}{c}{AR}&                
 \mc{1}{c}{S}&                 \mc{1}{c}{IA}&                \mc{1}{c}{EA}&                \mc{1}{c}{CA}&                
 \mc{1}{c}{IT}&                \mc{1}{c}{ET}&                \mc{1}{c}{E}&                 \mc{1}{c}{E}&                 
 \mc{1}{c}{I}&                 \mc{1}{c}{PA}&                \mc{1}{c}{P}&                 \mc{1}{c}{E}&                 
 \mc{1}{c}{I}&                 \mc{1}{c}{E}&                 \mc{1}{c}{C}&                 \mc{1}{c}{P}&                 
 \mc{1}{c}{HV}&                \mc{1}{c}{R}&                 \mc{1}{c}{LON}&               \mc{1}{c}{LAT}&               
 \mc{1}{c}{D$_{\rm Jup}$}&     \mc{1}{c}{ROT}&               \mc{1}{c}{S$_{\rm LON}$}&     \mc{1}{c}{S$_{\rm LAT}$}&     
 \mc{1}{c}{V}&                 \mc{1}{c}{VEF}&               \mc{1}{c}{M}&                 \mc{1}{c}{MEF}\\
\end{tabular*} \end{sidewaystable}                                                     
\clearpage                                                                      
\begin{sidewaystable} 

\tiny
\hspace{-4cm}
\begin{tabular*}{26cm} {@{\extracolsep{\fill}}                                  
rrrrrrrrrrrrrrrcrrrrcrrrrrrrrrrrrrrrrrrrrr}                                     
 \mc{1}{c}{No.}&               \mc{1}{c}{IMP. DATE}&         \mc{1}{c}{C}&                 \mc{1}{c}{AR}&                
 \mc{1}{c}{S}&                 \mc{1}{c}{IA}&                \mc{1}{c}{EA}&                \mc{1}{c}{CA}&                
 \mc{1}{c}{IT}&                \mc{1}{c}{ET}&                \mc{1}{c}{E}&                 \mc{1}{c}{E}&                 
 \mc{1}{c}{I}&                 \mc{1}{c}{PA}&                \mc{1}{c}{P}&                 \mc{1}{c}{E}&                 
 \mc{1}{c}{I}&                 \mc{1}{c}{E}&                 \mc{1}{c}{C}&                 \mc{1}{c}{P}&                 
 \mc{1}{c}{HV}&                \mc{1}{c}{R}&                 \mc{1}{c}{LON}&               \mc{1}{c}{LAT}&               
 \mc{1}{c}{D$_{\rm Jup}$}&     \mc{1}{c}{ROT}&               \mc{1}{c}{S$_{\rm LON}$}&     \mc{1}{c}{S$_{\rm LAT}$}&     
 \mc{1}{c}{V}&                 \mc{1}{c}{VEF}&               \mc{1}{c}{M}&                 \mc{1}{c}{MEF}\\
\end{tabular*} \end{sidewaystable}                                                     
\clearpage                                                                      
\begin{sidewaystable} 
\tiny	
\hspace{-4cm}
\begin{tabular*}{26cm} {@{\extracolsep{\fill}}                                  
rrrrrrrrrrrrrrrcrrrrcrrrrrrrrrrrrrrrrrrrrr}                                     
 \mc{1}{c}{No.}&               \mc{1}{c}{IMP. DATE}&         \mc{1}{c}{C}&                 \mc{1}{c}{AR}&                
 \mc{1}{c}{S}&                 \mc{1}{c}{IA}&                \mc{1}{c}{EA}&                \mc{1}{c}{CA}&                
 \mc{1}{c}{IT}&                \mc{1}{c}{ET}&                \mc{1}{c}{E}&                 \mc{1}{c}{E}&                 
 \mc{1}{c}{I}&                 \mc{1}{c}{PA}&                \mc{1}{c}{P}&                 \mc{1}{c}{E}&                 
 \mc{1}{c}{I}&                 \mc{1}{c}{E}&                 \mc{1}{c}{C}&                 \mc{1}{c}{P}&                 
 \mc{1}{c}{HV}&                \mc{1}{c}{R}&                 \mc{1}{c}{LON}&               \mc{1}{c}{LAT}&               
 \mc{1}{c}{D$_{\rm Jup}$}&     \mc{1}{c}{ROT}&               \mc{1}{c}{S$_{\rm LON}$}&     \mc{1}{c}{S$_{\rm LAT}$}&     
 \mc{1}{c}{V}&                 \mc{1}{c}{VEF}&               \mc{1}{c}{M}&                 \mc{1}{c}{MEF}\\
\end{tabular*} \end{sidewaystable}                                                     
\clearpage                                                                      
\begin{sidewaystable}
\tiny
\hspace{-4cm}
\begin{tabular*}{26cm} {@{\extracolsep{\fill}}                                  
rrrrrrrrrrrrrrrcrrrrcrrrrrrrrrrrrrrrrrrrrr}                                     
 \mc{1}{c}{No.}&               \mc{1}{c}{IMP. DATE}&         \mc{1}{c}{C}&                 \mc{1}{c}{AR}&                
 \mc{1}{c}{S}&                 \mc{1}{c}{IA}&                \mc{1}{c}{EA}&                \mc{1}{c}{CA}&                
 \mc{1}{c}{IT}&                \mc{1}{c}{ET}&                \mc{1}{c}{E}&                 \mc{1}{c}{E}&                 
 \mc{1}{c}{I}&                 \mc{1}{c}{PA}&                \mc{1}{c}{P}&                 \mc{1}{c}{E}&                 
 \mc{1}{c}{I}&                 \mc{1}{c}{E}&                 \mc{1}{c}{C}&                 \mc{1}{c}{P}&                 
 \mc{1}{c}{HV}&                \mc{1}{c}{R}&                 \mc{1}{c}{LON}&               \mc{1}{c}{LAT}&               
 \mc{1}{c}{D$_{\rm Jup}$}&     \mc{1}{c}{ROT}&               \mc{1}{c}{S$_{\rm LON}$}&     \mc{1}{c}{S$_{\rm LAT}$}&     
 \mc{1}{c}{V}&                 \mc{1}{c}{VEF}&               \mc{1}{c}{M}&                 \mc{1}{c}{MEF}\\
\end{tabular*} \end{sidewaystable}                                                     
\clearpage                                                                      
\begin{sidewaystable}
\tiny
\hspace{-4cm}
\begin{tabular*}{26cm} {@{\extracolsep{\fill}}                                  
rrrrrrrrrrrrrrrcrrrrcrrrrrrrrrrrrrrrrrrrrr}                                     
 \mc{1}{c}{No.}&               \mc{1}{c}{IMP. DATE}&         \mc{1}{c}{C}&                 \mc{1}{c}{AR}&                
 \mc{1}{c}{S}&                 \mc{1}{c}{IA}&                \mc{1}{c}{EA}&                \mc{1}{c}{CA}&                
 \mc{1}{c}{IT}&                \mc{1}{c}{ET}&                \mc{1}{c}{E}&                 \mc{1}{c}{E}&                 
 \mc{1}{c}{I}&                 \mc{1}{c}{PA}&                \mc{1}{c}{P}&                 \mc{1}{c}{E}&                 
 \mc{1}{c}{I}&                 \mc{1}{c}{E}&                 \mc{1}{c}{C}&                 \mc{1}{c}{P}&                 
 \mc{1}{c}{HV}&                \mc{1}{c}{R}&                 \mc{1}{c}{LON}&               \mc{1}{c}{LAT}&               
 \mc{1}{c}{D$_{\rm Jup}$}&     \mc{1}{c}{ROT}&               \mc{1}{c}{S$_{\rm LON}$}&     \mc{1}{c}{S$_{\rm LAT}$}&     
 \mc{1}{c}{V}&                 \mc{1}{c}{VEF}&               \mc{1}{c}{M}&                 \mc{1}{c}{MEF}\\
\end{tabular*} \end{sidewaystable}                                                     
\clearpage                                                                      
\begin{sidewaystable}
\tiny
\hspace{-4cm}
\begin{tabular*}{26cm} {@{\extracolsep{\fill}}                                  
rrrrrrrrrrrrrrrcrrrrcrrrrrrrrrrrrrrrrrrrrr}                                     
 \mc{1}{c}{No.}&               \mc{1}{c}{IMP. DATE}&         \mc{1}{c}{C}&                 \mc{1}{c}{AR}&                
 \mc{1}{c}{S}&                 \mc{1}{c}{IA}&                \mc{1}{c}{EA}&                \mc{1}{c}{CA}&                
 \mc{1}{c}{IT}&                \mc{1}{c}{ET}&                \mc{1}{c}{E}&                 \mc{1}{c}{E}&                 
 \mc{1}{c}{I}&                 \mc{1}{c}{PA}&                \mc{1}{c}{P}&                 \mc{1}{c}{E}&                 
 \mc{1}{c}{I}&                 \mc{1}{c}{E}&                 \mc{1}{c}{C}&                 \mc{1}{c}{P}&                 
 \mc{1}{c}{HV}&                \mc{1}{c}{R}&                 \mc{1}{c}{LON}&               \mc{1}{c}{LAT}&               
 \mc{1}{c}{D$_{\rm Jup}$}&     \mc{1}{c}{ROT}&               \mc{1}{c}{S$_{\rm LON}$}&     \mc{1}{c}{S$_{\rm LAT}$}&     
 \mc{1}{c}{V}&                 \mc{1}{c}{VEF}&               \mc{1}{c}{M}&                 \mc{1}{c}{MEF}\\
\end{tabular*} \end{sidewaystable}                                                     
\clearpage                                                                      
\begin{sidewaystable}
\tiny
\hspace{-4cm}
\begin{tabular*}{26cm} {@{\extracolsep{\fill}}                                  
rrrrrrrrrrrrrrrcrrrrcrrrrrrrrrrrrrrrrrrrrr}                                     
 \mc{1}{c}{No.}&               \mc{1}{c}{IMP. DATE}&         \mc{1}{c}{C}&                 \mc{1}{c}{AR}&                
 \mc{1}{c}{S}&                 \mc{1}{c}{IA}&                \mc{1}{c}{EA}&                \mc{1}{c}{CA}&                
 \mc{1}{c}{IT}&                \mc{1}{c}{ET}&                \mc{1}{c}{E}&                 \mc{1}{c}{E}&                 
 \mc{1}{c}{I}&                 \mc{1}{c}{PA}&                \mc{1}{c}{P}&                 \mc{1}{c}{E}&                 
 \mc{1}{c}{I}&                 \mc{1}{c}{E}&                 \mc{1}{c}{C}&                 \mc{1}{c}{P}&                 
 \mc{1}{c}{HV}&                \mc{1}{c}{R}&                 \mc{1}{c}{LON}&               \mc{1}{c}{LAT}&               
 \mc{1}{c}{D$_{\rm Jup}$}&     \mc{1}{c}{ROT}&               \mc{1}{c}{S$_{\rm LON}$}&     \mc{1}{c}{S$_{\rm LAT}$}&     
 \mc{1}{c}{V}&                 \mc{1}{c}{VEF}&               \mc{1}{c}{M}&                 \mc{1}{c}{MEF}\\
\end{tabular*} \end{sidewaystable}                                                     
\clearpage                                                                      
\begin{sidewaystable}
\tiny
\hspace{-4cm}
\begin{tabular*}{26cm} {@{\extracolsep{\fill}}                                  
rrrrrrrrrrrrrrrcrrrrcrrrrrrrrrrrrrrrrrrrrr}                                     
 \mc{1}{c}{No.}&               \mc{1}{c}{IMP. DATE}&         \mc{1}{c}{C}&                 \mc{1}{c}{AR}&                
 \mc{1}{c}{S}&                 \mc{1}{c}{IA}&                \mc{1}{c}{EA}&                \mc{1}{c}{CA}&                
 \mc{1}{c}{IT}&                \mc{1}{c}{ET}&                \mc{1}{c}{E}&                 \mc{1}{c}{E}&                 
 \mc{1}{c}{I}&                 \mc{1}{c}{PA}&                \mc{1}{c}{P}&                 \mc{1}{c}{E}&                 
 \mc{1}{c}{I}&                 \mc{1}{c}{E}&                 \mc{1}{c}{C}&                 \mc{1}{c}{P}&                 
 \mc{1}{c}{HV}&                \mc{1}{c}{R}&                 \mc{1}{c}{LON}&               \mc{1}{c}{LAT}&               
 \mc{1}{c}{D$_{\rm Jup}$}&     \mc{1}{c}{ROT}&               \mc{1}{c}{S$_{\rm LON}$}&     \mc{1}{c}{S$_{\rm LAT}$}&     
 \mc{1}{c}{V}&                 \mc{1}{c}{VEF}&               \mc{1}{c}{M}&                 \mc{1}{c}{MEF}\\
\end{tabular*} \end{sidewaystable}                                                     
\clearpage                                                                      
\begin{sidewaystable}
\tiny
\hspace{-4cm}
\begin{tabular*}{26cm} {@{\extracolsep{\fill}}                                  
rrrrrrrrrrrrrrrcrrrrcrrrrrrrrrrrrrrrrrrrrr}                                     
 \mc{1}{c}{No.}&               \mc{1}{c}{IMP. DATE}&         \mc{1}{c}{C}&                 \mc{1}{c}{AR}&                
 \mc{1}{c}{S}&                 \mc{1}{c}{IA}&                \mc{1}{c}{EA}&                \mc{1}{c}{CA}&                
 \mc{1}{c}{IT}&                \mc{1}{c}{ET}&                \mc{1}{c}{E}&                 \mc{1}{c}{E}&                 
 \mc{1}{c}{I}&                 \mc{1}{c}{PA}&                \mc{1}{c}{P}&                 \mc{1}{c}{E}&                 
 \mc{1}{c}{I}&                 \mc{1}{c}{E}&                 \mc{1}{c}{C}&                 \mc{1}{c}{P}&                 
 \mc{1}{c}{HV}&                \mc{1}{c}{R}&                 \mc{1}{c}{LON}&               \mc{1}{c}{LAT}&               
 \mc{1}{c}{D$_{\rm Jup}$}&     \mc{1}{c}{ROT}&               \mc{1}{c}{S$_{\rm LON}$}&     \mc{1}{c}{S$_{\rm LAT}$}&     
 \mc{1}{c}{V}&                 \mc{1}{c}{VEF}&               \mc{1}{c}{M}&                 \mc{1}{c}{MEF}\\
\end{tabular*} \end{sidewaystable}                                                     
\clearpage                                                                      
\begin{sidewaystable}
\tiny
\hspace{-4cm}
\begin{tabular*}{26cm} {@{\extracolsep{\fill}}                                  
rrrrrrrrrrrrrrrcrrrrcrrrrrrrrrrrrrrrrrrrrr}                                     
 \mc{1}{c}{No.}&               \mc{1}{c}{IMP. DATE}&         \mc{1}{c}{C}&                 \mc{1}{c}{AR}&                
 \mc{1}{c}{S}&                 \mc{1}{c}{IA}&                \mc{1}{c}{EA}&                \mc{1}{c}{CA}&                
 \mc{1}{c}{IT}&                \mc{1}{c}{ET}&                \mc{1}{c}{E}&                 \mc{1}{c}{E}&                 
 \mc{1}{c}{I}&                 \mc{1}{c}{PA}&                \mc{1}{c}{P}&                 \mc{1}{c}{E}&                 
 \mc{1}{c}{I}&                 \mc{1}{c}{E}&                 \mc{1}{c}{C}&                 \mc{1}{c}{P}&                 
 \mc{1}{c}{HV}&                \mc{1}{c}{R}&                 \mc{1}{c}{LON}&               \mc{1}{c}{LAT}&               
 \mc{1}{c}{D$_{\rm Jup}$}&     \mc{1}{c}{ROT}&               \mc{1}{c}{S$_{\rm LON}$}&     \mc{1}{c}{S$_{\rm LAT}$}&     
 \mc{1}{c}{V}&                 \mc{1}{c}{VEF}&               \mc{1}{c}{M}&                 \mc{1}{c}{MEF}\\
\end{tabular*} \end{sidewaystable} 

\clearpage
\pagestyle{plain}

\end{document}